\documentstyle[referee,epsf]{mn}
\input psfig.tex
\def\lsim{\lower.5ex\hbox{$\; \buildrel < \over \sim \;$}}
\def\gsim{\lower.5ex\hbox{$\; \buildrel > \over \sim \;$}}
\def \simeq{\lower.3ex\hbox{$\; \buildrel \sim \over - \;$}}
\def\ch{\lower-0.55ex\hbox{--}\kern-0.55em{\lower0.15ex\hbox{$h$}}}
\def\lh{\lower-0.55ex\hbox{--}\kern-0.55em{\lower0.15ex\hbox{$\lambda$}}}
\newif\ifAMStwofonts
\ifoldfss
  \ifCUPmtlplainloaded \else
    \NewTextAlphabet{textbfit} {cmbxti10} {}
    \NewTextAlphabet{textbfss} {cmssbx10} {}
    \NewMathAlphabet{mathbfit} {cmbxti10} {} 
    \NewMathAlphabet{mathbfss} {cmssbx10} {} 
  \fi
  \ifAMStwofonts
    \ifCUPmtlplainloaded \else
      \NewSymbolFont{upmath} {eurm10}
      \NewSymbolFont{AMSa} {msam10}
      \NewMathSymbol{\upi}     {0}{upmath}{19}
      \NewMathSymbol{\umu}     {0}{upmath}{16}
      \NewMathSymbol{\upartial}{0}{upmath}{40}
      \NewMathSymbol{\leqslant}{3}{AMSa}{36}
      \NewMathSymbol{\geqslant}{3}{AMSa}{3E}

    \fi
  \fi
\fi 
\ifnfssone
  \newmathalphabet{\mathit}
  \addtoversion{normal}{\mathit}{cmr}{m}{it}
  \addtoversion{bold}{\mathit}{cmr}{bx}{it}
  \newmathalphabet{\mathbfit} 
  \addtoversion{normal}{\mathbfit}{cmr}{bx}{it}
  \addtoversion{bold}{\mathbfit}{cmr}{bx}{it}
  \newmathalphabet{\mathbfss} 
  \addtoversion{normal}{\mathbfss}{cmss}{bx}{n}
  \addtoversion{bold}{\mathbfss}{cmss}{bx}{n}
  \ifAMStwofonts
    \ifCUPmtlplainloaded \else
      \UseAMStwoboldmath
      \makeatletter
      \new@mathgroup\upmath@group
      \define@mathgroup\mv@normal\upmath@group{eur}{m}{n}
      \define@mathgroup\mv@bold\upmath@group{eur}{b}{n}
      \edef\UPM{\hexnumber\upmath@group}
      \new@mathgroup\amsa@group
      \define@mathgroup\mv@normal\amsa@group{msa}{m}{n}
      \define@mathgroup\mv@bold\amsa@group{msa}{m}{n}
      \edef\AMSa{\hexnumber\amsa@group}
      \makeatother
      \mathchardef\upi="0\UPM19
      \mathchardef\umu="0\UPM16
      \mathchardef\upartial="0\UPM40
      \mathchardef\leqslant="3\AMSa36
      \mathchardef\geqslant="3\AMSa3E
    \fi
  \fi
\fi 

\ifnfsstwo
  \DeclareMathAlphabet{\mathbfit}{OT1}{cmr}{bx}{it}
  \SetMathAlphabet\mathbfit{bold}{OT1}{cmr}{bx}{it}
  \DeclareMathAlphabet{\mathbfss}{OT1}{cmss}{bx}{n}
  \SetMathAlphabet\mathbfss{bold}{OT1}{cmss}{bx}{n}
  \ifAMStwofonts
    \ifCUPmtlplainloaded \else
      \SetSymbolFont{UPM}{bold}{U}{eur}{b}{n}
      \DeclareSymbolFont{AMSa}{U}{msa}{m}{n}
      \DeclareMathSymbol{\upi}{0}{UPM}{"19}
      \DeclareMathSymbol{\umu}{0}{UPM}{"16}
      \DeclareMathSymbol{\upartial}{0}{UPM}{"40}
      \DeclareMathSymbol{\leqslant}{3}{AMSa}{"36}
      \DeclareMathSymbol{\geqslant}{3}{AMSa}{"3E}
    \fi
  \fi
\fi 

\ifCUPmtlplainloaded \else
  \ifAMStwofonts \else 
    \def\upi{\pi}
    \def\umu{\mu}
    \def\upartial{\partial}
  \fi
\fi

\title{Properties of Accretion Shock Waves in Viscous Flows Around Black Holes}

\author[Sandip K. Chakrabarti and Santabrata Das]
       {Sandip K. Chakrabarti$^{1,2}$ and Santabrata Das$^{1}$\\
$^1$ S.N. Bose National Centre for Basic Sciences,\\
JD-Block, Sector III, Salt Lake, Calcutta 700098, India\\
$^2$ Also at Centre for Space Physics, Chalantika 43, Garia Station Rd., Garia, Kolkata 700084, India\\} 

\begin{document}

\maketitle

\begin{abstract}

Accretion flows having low angular momentum and low viscosity can have
standing shock waves. These shocks arise due to the presence of multiple
sonic points in the flow. We study the region of the parameter space 
in which multiple sonic points occur in viscous flows in the absence of cooling. We also separate
the parameter space in regions allowing steady shocks and oscillating shocks. 
We quantify the nature of two critical viscosities which separate the 
flow topologies. A post-shock region being hotter, it emits harder X-rays
and oscillating shocks cause oscillating X-ray intensities giving rise to
quasi-periodic oscillations. We show that with 
the increase in viscosity parameter, the shock always 
moves closer to the black hole. This implies an enhancement of 
the quasi-periodic oscillation frequency as viscosity is increased.

\end{abstract}

\noindent MNRAS (in press)

\section{Introduction}

In the standard theory of thin accretion flows around black holes (Shakura
\& Sunyaev, 1973, hereafter referred to as SS73) viscosity plays a major role. 
Viscosity transports angular momentum outwards and allows matter to sink 
into the potential well formed by the central compact object. In this model, the flow angular momentum is assumed 
to be Keplerian and this is the standard notion about how matter is  accreted. 
However, Chakrabarti \& Molteni (1995, hereafter referred to as Paper I), 
and Lanzafame, Molteni \& Chakrabarti (1998, hereafter referred to as Paper II), 
through extensive numerical simulations showed that the angular momentum 
distribution depends strictly on the viscosity parameter and the way
the viscous stress is defined. They showed that close to a black hole, 
the disk does not have a Keplerian distribution. This is because the flow must be supersonic 
on the horizon (Chakrabarti, 1990a) whereas a Keplerian disk is always subsonic
(SS73). In Papers I  and II, it was shown that for a large region of the
parameter space, shocks may form in accretion flows and when viscosity is increased
beyond a critical value (Chakrabarti, 1990ab; Chakrabarti 1996, hereafter C96a), the shocks disappear.

Paper I also improved the concept of viscosity parameter $\alpha$ (SS73):
it argued that in a generalized flow with significant radial velocity $\vartheta$, 
the viscous stress $w_{\phi r}$ should not be equated to $-\alpha P$ as 
in SS73, where $P$ is the total pressure, but to $-\alpha_\Pi (P+\rho \vartheta^2)$,
(actually, its vertically integrated value using a thin disk approximation)
where, $\rho$ is the density and a subscript $\Pi$ is given to $\alpha$ to distinguish it
from the Shakura-Sunyaev viscosity parameter. The latter prescription naturally 
goes over to the original prescription when radial velocity is unimportant as in the case 
of a standard Keplerian disk model (SS73), however, when the radial 
velocity is important as in the transonic flow solutions (Chakrabarti 1990a),
the latter definition preserves the angular momentum even across axisymmetric 
discontinuities, such as accretion shocks. The reason is, according to 
the Rankine-Hugoniot conditions (Landau \& Lifshitz, 1959), in a steady flow, the sum 
of thermal pressure and ram pressure, i.e., $P+\rho \vartheta^2$ 
is continuous across discontinuities. This makes the viscous stress $w_{r\phi}$ 
continuous across axisymmetric discontinuities as well.

In an earlier study, Chakrabarti (1989a, hereafter C89a) considered the transonic properties 
of isothermal accretion flows and showed that for a large region of the 
parameter space spanned by the specific angular momentum and the temperature of the
flow, an accretion disk can have standing shock waves. The specific angular 
momentum of the disk was smaller than that of a Keplerian disk everywhere. This flows come about 
especially when the matter is accreted from the winds of a binary companion. 
Subsequently, Chakrabarti (1990b, hereafter C90b) showed that inclusion of viscosity
reduces the region of the parameter space in that, 
at a sufficiently high viscosity, the Rankine-Hugoniot conditions which 
must be satisfied at a steady shock are not satisfied anywhere in the flow. 
Existence of standing shocks in sub-Keplerian inviscid accretion disks have been tested independently 
by several groups since then (Nobuta and Hanawa 1994; Yang and Kafatos, 1995; 
Lu and Yuan, 1997). Numerical simulations have also been carried out with 
several independent codes such as Smootherd Particle Hydrodynamics (SPH) and Total Variation Diminishing
(TVD) and distinct standing shocks were found exactly at the predicted locations 
(Chakrabarti \& Molteni, 1993; Molteni, Ryu \& Chakrabarti, 1996).

In more recent years, it has become evident that the standing shocks
may be very important in explaining the spectral properties of 
black hole candidates (Chakrabarti \& Titarchuk, 1995, hereafter CT95) as the
post-shock region behaves as the boundary layer where accreting 
matter dissipates its thermal energy and generates hard X-ray
by inverse Comptonization. C96a considered unification of solutions
of winds and accretion around compact objects. However, the cooling was
treated in terms of a parameter and no parameter space was studied. 
The post-shock region is also found to be responsible to 
produce relativistic outflows (Chakrabarti, 1999; Chattopadhyay and Chakrabarti, 2002). 
Furthermore, numerical simulations indicated that 
the shocks may be oscillating at nearby regions of the parameter space 
in presence of cooling effects (Molteni, Sponholz \& Chakrabarti, 1996) 
and the shock oscillations correctly 
explain intricate properties of quasi-periodic oscillations (Chakrabarti \& 
Manickam, 2000). Recent observations do support the presence of 
sub-Keplerian flows in accretion disks (Smith et al. 2001; Smith, Heindl and Swank, 2002).

In view of the importance of the sub-Keplerian flows we plan to re-investigate the
work done on isothermal flow by C89a and C90b by extending them to study {\it
polytropic flows} to check the properties of shock waves in viscous flows. What is more, 
unlike C89a and C90b, we investigate the behaviour of the solutions
in the entire parameter space spanned by the specific energy, angular 
momentum and the viscosity. In C96a some work was done, the parameter space was not explored.
We find very important results: even when 
the viscosity parameter is very high, the flow continues to have three sonic points: 
a prime condition to have a standing or oscillating shock waves.
However, the parameter space for standing shock waves is gradually reduced
with the increase of viscosity. On the other hand, we discover that the 
shock location itself is reduced with the increase in viscosity parameter.

We wish to emphasize that the problem at hand is by no means a trivial extension 
of previous works. In an accretion flow, where the flow is subsonic at a large 
distance and is necessarily supersonic on the horizon, the flow has to first become 
supersonic at a sonic point and then after the shock transition where it becomes 
sub-sonic, the flow must again pass through the inner sonic point before entering 
into the black hole. In studying flows with constant energy (Chakrabarti, 1989b, hereafter C89b) 
or an isothermal flows (C89a), both the sonic points were known when the 
so-called `eigen-values', namely, the specific energy (for polytropic flow) or 
temperature (isothermal flow) and the specific angular momentum, are supplied. 
In the present situation, neither of these two quantities is constant in the flow
since the viscosity will heat up the gas, increase thermal energy and at the 
same time reduce the specific angular momentum as the flow proceeds towards the 
black hole. Thus, the inner sonic point, through which the flow will pass after 
the shock, is not known before the entire problem is actually solved. We have devised 
a novel way to solve the entire problem by iterating the location of the inner sonic 
point till the shock condition is satisfied. We have identified the topologies which
are essential for shock formation. We have also identified the parameter space
which will have solutions with three sonic points but need not have standing shocks.
These solutions generally produce oscillating shocks as shown by Ryu, Chakrabarti and Molteni, (1997).
In C96a, some effects of viscous heating were studied and cooling effect was
chosen to be proportional to the heating effect for simplicity. No parameter space study was made.
In the present paper, we ignore cooling completely. 
Exact effects of various cooling processes and their influence on the
parameter space will be discussed elsewhere (Das \& Chakrabarti, 2003).

The plan of the present paper is the following: in the next Section, we 
present the model equations. In \S 3, we present the sonic point analysis.
In \S 4, we  study the global solution topology. In \S 5, we classify the 
parameter space in terms of whether a global solution has triple sonic points
or not. In \S 6, we classify the region with triple sonic points further
to indicate which region may allow standing shocks and which region may allow
oscillating shocks in presence of viscosity. We showed in particular that matter 
with a very low angular momentum may allow shocks even when the viscosity parameter
is very high. In C96a it was shown that topologies are changed
with viscosity and there exists two critical viscosity parameters at which such changes take
place.  In \S 7, we quantify these critical viscosity parameters.
Finally, in \S 8, we discuss about the relevance of shock shocks in the 
context of quasi-periodic oscillations and make concluding remarks.

\section{Model Equations}

We consider a steady, thin, viscous, axisymmetric accretion 
flow on to a Schwarzschild black hole. The space-time 
geometry around a Schwarzschild black hole is described
by the pseudo-Newtonian potential introduced by Paczy\'nski \& Wiita (1980). 
Here, one uses the pseudo-Newtonian potential given by, 
$g(r) = -{\frac {GM}{r-2GM_{BH}/c^2}}$.
We consider the units of velocity, distance and time to be $c$, $r_g=2GM_{BH}/c^2$ 
and $2GM_{BH}/c^3$ respectively where $c$ is the velocity of light, 
$G$ and $M_{BH}$ are the gravitational constant and mass of the black hole respectively.
In this unit, defining $x=r/r_g$, we get the potential as $g(x)=-\frac{1}{2(x-1)}$.
We assume the disk to be in {\it hydrostatic} equilibrium in the vertical direction.

In the steady state, the dimensionless hydrodynamic equations
that govern the infalling matter are the followings (C96a).

\noindent (a) Radial Momentum equation :
$$
\vartheta \frac {d\vartheta}{dx}+\frac {1}{\rho}\frac {dP}{dx}
-\frac {\lambda(x)^2}{x^3}+\frac {1}{2(x-1)^2}=0
\eqno{(1a)}
$$

\noindent (b) Baryon number conservation equation :

$$
\dot M = \Sigma \vartheta x
\eqno{(1b)}
$$
apart from a geometric constant.

\noindent (c) Angular momentum conservation equation :

$$
\vartheta \frac {d\lambda(x)}{dx}+\frac{1}{\Sigma x}
\frac {d}{dx}\left( x^2 W_{x\phi}\right)=0,
\eqno{(1c)}
$$

and finally,

\noindent (d) The entropy generation equation :

$$
\Sigma \vartheta T \frac {ds}{dx}= Q^+ - Q^- .
\eqno{(1d)}
$$

The local variables $\vartheta$, $\rho$, $P$ and $\lambda(x)$ in the 
above equations are the radial velocity, density, isotropic
pressure and specific angular momentum of the flow
respectively. Here $\Sigma$ and $W_{x\phi}$ are the 
vertically integrated density (Matsumoto et al. 1984) and the viscous 
stress, $s$ is the entropy density of the flow, $T$ is the local temperature.
$Q^+$ and $Q^-$ are the heat gained and lost by the flow 
(integrated in the vertical direction) respectively. 

In our model of the disk which is assumed to be in hydrostatic equilibrium in the 
vertical direction, local disk height is obtained by
equating the pressure gradient force in the vertical direction with the
component of the gravitational force in that direction. 
The half thickness of the disk is obtained as:
$$
h=ax^{1/2}(x-1).
\eqno{(2)}
$$
Here, $a$ is the adiabatic sound speed defined as $a=\sqrt {\gamma P/\rho}$.
As discussed in the introduction, we shall use the viscosity prescription 
of Paper I valid rigorously for flows with significant radial motion.   
Thus the viscous stress is:
$$
W_{x\phi} = - \alpha_{\Pi}\Pi,
\eqno{(3)}
$$ 
where, $\Pi=W+\Sigma \vartheta^2$. As mentioned before, this will ensure 
that the viscous stress is continuous
across the axisymmetric shock wave that we are studying here.
It is to be noted that in SS73 prescription, $W_{x\phi}$ is not continuous
across the shock. Thus, the stress would transport angular momentum 
at different rates on two sides of the shock which would always
move the shock one way or the other. This is unphysical, since 
in the absence of viscosity, a standing, axisymmetric shock is perfectly stable.
It is impossible that an infinitesimal viscosity should 
destabilize the shock. However, this would have been the case if SS73 prescription 
were rigorously correct.

\section{Sonic Point Analysis}

At the outer edge of the accretion disk, matter has almost 
zero radial velocity even though it enters into the black hole with 
velocity of light $c$. Thus, during accretion, at some point,
matter velocity should exactly match with the sound speed.
This point is called a critical point or a sonic point. When 
matter crosses a sonic point, it becomes transonic. As Chakrabarti 
(C89ab, C90ab) pointed out, depending on the initial 
parameters, a flow may have multiple sonic points and therefore,
depending on whether the shock conditions are satisfied or not,
a flow may or may not have a standing shock.

For the sake of completeness, we carry out the 
sonic point analysis by solving the above Eqs. (1-3) 
using a method similar to that used in C89b. 

\subsection{Sonic point conditions}

In the present analysis, we use MISStress prescription (C96a)
for computing $Q^+$ and  $W_{x\phi}$ is obtained from Eq. (3).
For the accretion flow, the entropy equation (Eq. 4) can be simplified as,
$$
\frac {\vartheta}{\gamma -1}\left[ \frac {1}{\rho}
\frac {dP}{dx}-\frac {\gamma P}{\rho^2}\frac {d\rho}{dx}\right]
=\frac {Q^--Q^+}{\rho h}=C-H ,
\eqno{(4)}
$$
and then $H (= Q^+/\rho h)$ takes the form,
$$
H = A x (ga^2+\gamma\vartheta^2)\frac {d\Omega}{dx},
\eqno{(5)}
$$ 
where, $\gamma$ is the adiabatic index, $A = -\alpha_{\Pi}\frac {I_n}
{\gamma}$ and $g = \frac {I_{n+1}}{I_n}$. Here, $\Omega(x)$ is the 
angular velocity of the accreting matter at the radial 
distance $x$, $n$ is the polytropic index $(n= \frac {1}{\gamma-1})$, 
$I_n$ and $I_{n+1}$ come from the definition of the vertically averaged density and 
pressure (Matsumoto et al. 1984). 

In the present analysis, we use $Q_-=0$, ${\it i.e.}$, the cooling is ignored. This
would be strictly valid if the accretion rate is low, so that the
loss of energy by bremsstrahlung cooling is insignificant compared to the
rest mass energy.

After some simple algebra and eliminating $da/dr$ etc.
we get from the governing equations Eqs. 1(a-c) and Eq. (2) the
following first order linear differential equation:
$$
\frac {d\vartheta}{dx}=\frac {N}{D} ,
\eqno{(6)}
$$
where, the numerator $N$ is, 
$$
N =-\frac {\alpha_\Pi A (a^2g+\gamma \vartheta^2)^2}{\gamma x}
-\left[ \frac {\lambda^2}{x^3}-\frac {1}{2(x-1)^2}\right]
\left[ 2\alpha_\Pi g A (a^2g+\gamma\vartheta^2)+ \frac {(\gamma+1)
\vartheta^2}{(\gamma-1)} \right]
$$
$$
-\frac {\vartheta^2 a^2(5x-3)}{x(\gamma-1)(x-1)}
-\frac {\alpha_\Pi g A a^2(5x-3)(a^2g+\gamma\vartheta^2)}{\gamma x(x-1)}
+\frac {2\lambda A \vartheta (a^2g+\gamma\vartheta^2)}{x^2}
\eqno(7)
$$
and the denominator $D$ is
$$
D = \frac {2a^2\vartheta}{(\gamma-1)}-\frac {(\gamma+1)\vartheta^3} {(\gamma-1)}
-A\alpha_\Pi\vartheta(a^2g+\gamma\vartheta^2)
\left[ (2g-1)-\frac {a^2g}{\gamma\vartheta^2}\right] .
\eqno(8)
$$
Both $N$ and $D$ are algebraic equations which makes this model easily tractable.

At the sonic point, both the numerator and the denominator must 
vanish simultaneously. For $D=0$, one can get the 
expression for the Mach Number $M(x_c)$ at the sonic point and is given by,
$$
M(x_c) =\sqrt {\frac{-m_b - \sqrt{m^2_b-4m_a m_c}}{2m_a}}
$$
where,
$$ 
m_a=-A\alpha_\Pi \gamma^2(\gamma-1)(2g-1)-\gamma(\gamma+1),
$$
$$
m_b=2\gamma-2A\alpha_\pi g\gamma(\gamma-1)(g-1)
$$ 
$$
m_c=A\alpha_\pi g^2 (\gamma-1).
$$ 
In the weak viscosity limit, $\alpha_\Pi \rightarrow 0$ and 
the Mach number at the sonic point 
$M(x_c) \approx \sqrt {\frac {2}{\gamma +1}} \ \ {\rm for} \ \ \alpha_{\Pi} \rightarrow 0 $
a result obtained in C89b.

Setting $N = 0$, we get an algebraic equation for the sound speed at the 
sonic point which is given by, 
$$
F({\cal E}_c, \lambda_c, x_c)=-\left[ \frac {\alpha_\Pi A \{g+\gamma M^2\}^2}
{\gamma x}+\frac {\alpha_\Pi A (5x-3)\{g+\gamma M^2\}}
{\gamma x(x-1)}+\frac {M^2(5x-3)}{x(\gamma-1)(x-1)} \right]a^2 
$$
$$
+\frac {2\lambda A M (g+\gamma M^2)}{x^2}a
-\left[ \frac {\lambda^2}{x^3}-\frac {1}{2(x-1)^2}\right]
\left[ 2\alpha_\Pi g A (g+\gamma M^2)+ \frac {(\gamma+1)
M^2}{(\gamma-1)} \right] =0
\eqno(9)
$$

We solve the above quadratic equation to obtain the sound speed at the sonic point.
Das et al. (2001) suggested that depending on a given set of initial parameters
accretion flow may have a maximum of four sonic points where one of the sonic points always lies
inside the black hole horizon for non-dissipative accretion flow. In our present study, 
we also expect a similar result.

\subsection{Nature of the sonic points}

A black hole accretion is always transonic. Thus the originally subsonic 
matter definitely has to pass through the sonic point to become supersonic 
before entering into the black hole. Depending on the initial parameters, a flow may have
multiple sonic points. Nature of sonic point depends on the value
of velocity gradients at the sonic point. It is easy to show that 
$\frac {d\vartheta}{dx}$ assumes two values at the sonic point.
One of them is valid for the accretion flow and the other is valid for the wind. If both the derivatives 
are real and of opposite signs, the sonic point is saddle 
type. When the derivatives are real and of the same sign, the sonic point is nodal
type. When the derivative is complex, the sonic point is spiral type 
(or O-type, for non dissipative flow). See, C90b for details of classifications. In order to 
form a standing shock, the flow must have more than one saddle type sonic points.

\begin {figure}
\vbox{
\centerline{
\psfig{figure=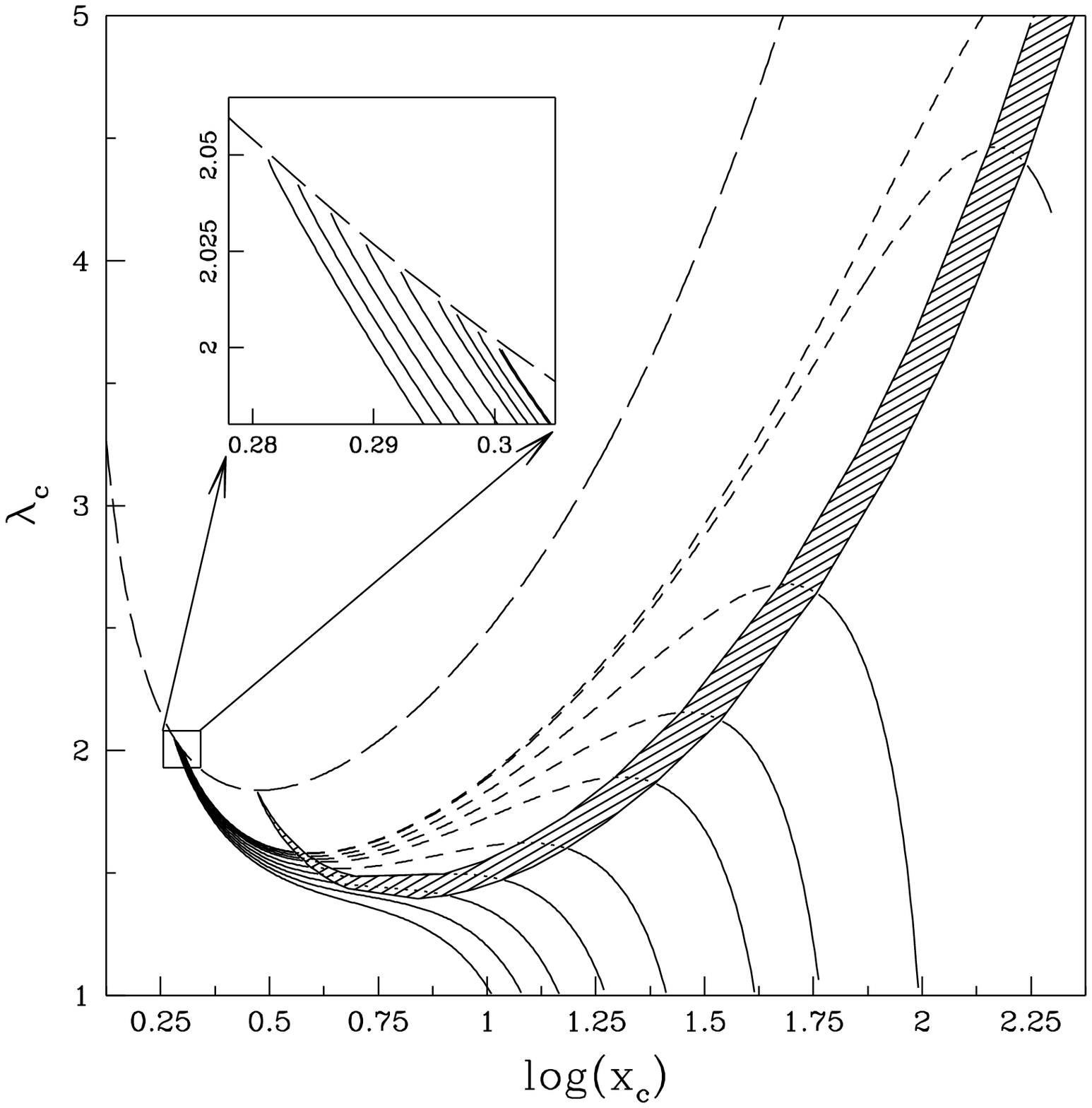,height=10truecm,width=8truecm}}}
\noindent{\small {\bf Fig. 1a:} Variation of specific angular momentum ($\lambda_{c}$)
as a function of the logarithmic sonic point location $(x_c)$ for the viscosity
parameter ($\alpha_\Pi = 0.1$). Long-dashed curve is the Keplerian
angular momentum distribution. Solid curves represent the saddle type sonic points, 
dotted curves represent the nodal type sonic points and the short-dashed curves are 
for the spiral type sonic points. Shaded area is the nodal type sonic point region.}
\end{figure}

In Figure 1a, we plot the variation of specific angular momentum ($\lambda_{c}$) 
as a function of the logarithmic sonic point location $(x_c)$ for a given viscosity
parameter ($\alpha_\Pi = 0.1$). Here different curves are drawn for different
specific energies at the sonic points. The energies, from the top curve to the bottom,
are given by: ${\cal E}_c = 0.0007, 0.001, 0.003, 0.005, 0.007, 0.011, 
0.015, 0.019, 0.023$ and $0.027$ respectively. The 
long-dashed curve at the top represents the Keplerian 
angular momentum distribution which is completely independent of the initial flow
parameters and depends only on the geometry. Solid part of the curves
represents the saddle type sonic points, dotted part of the curves represents 
the nodal type sonic points and the short-dashed part of the  curves are 
for the spiral type sonic points. First notice that
the sonic points always occur at angular momentum below Keplerian value. Notice that for
lower values of specific energy at the sonic point, an accretion flow contains
all the three types of sonic points in a systematic order: saddle --- nodal --- 
spiral --- nodal --- saddle for monotonic increase of location of sonic points. 
With the increase of energy ${\cal E}_c$ the region of spiral type sonic points 
gradually decreases and finally replaced by the nodal type sonic 
points though multiple sonic points still exist. Shaded area separates 
the nodal type sonic point region in the $\lambda_c - x_c$ plane.
With further increase of energy all the nodal type sonic points also
disappear and  are replaced by saddle type sonic points. In this case, the flow 
has only one sonic point for a given sub-Keplerian angular momentum. 
Thus, for a given angular momentum of the flow, there exists a range
of energy ${\cal E}_{min} < {{\cal E}_c} < {\cal E}_{max}$ such that the
flow has multiple sonic points. In the {\it inset}, we zoom a small
portion of the curve close the Keplerian value to  highlight the fact 
that the angular momentum at the sonic point always remains 
sub-Keplerian when cooling process is ignored.  
In future (Das \& Chakrabarti 2003, in preparation), we shall show that 
a flow can also be super-Keplerian when cooling is added.

\begin {figure}
\vbox{
\centerline{
\psfig{figure=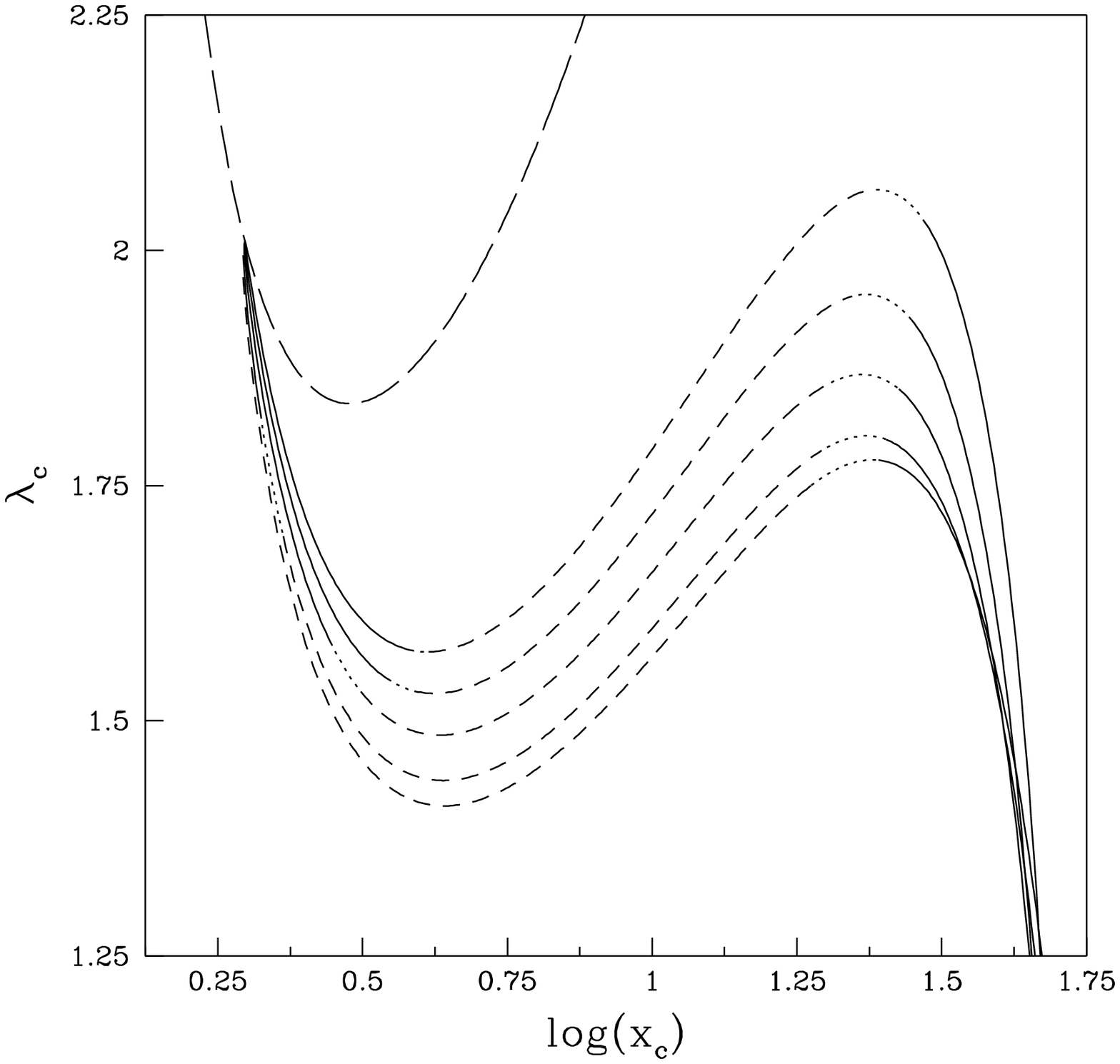,height=10truecm,width=8truecm}}}
\noindent{\small {\bf Fig. 1b:} 
Variation of angular momentum at the sonic point as the viscosity parameter is varied. Specific energy
at the inner sonic point is held fixed at $0.006$. From the uppermost to the lowermost curve: $\alpha_\Pi=0,
0.2,0.4,0.6$ and $0.7$ respectively. Other notations are the same as in Fig. 1a.}
\end{figure}

In Fig. 1b, we show a very important aspect of a viscous transonic flow. Here we 
show how the angular momentum at the sonic point varies when viscosity parameter $\alpha_\Pi$ 
is increased.  We hold the energy at the sonic point to be fixed at ${\cal E}_{in} = 0.006$. 
In the absence of viscosity ($\alpha_\Pi=0$, the uppermost curve), 
the flow has all the three types of sonic points. 
Similar to Fig. 1a, here also we indicated the saddle, nodal and spiral type sonic points by the 
solid, dotted and short-dashed curves respectively. The uppermost long dashed curve
represents the Keplerian angular momentum distribution.
With the increase of $\alpha_\Pi$, more
and more inner saddle type sonic points are replaced by nodal type sonic 
points and similarly nodal type are also replaced by spiral type sonic points. 
The curves, from the uppermost one to the lowermost one, are, for $\alpha_\Pi=0, 
\ 0.2, \ 0.4, \ 0.6, \ 0.7$ respectively.
For $\alpha_\Pi = 0.7$, all the inner saddle type sonic points disappear and
only the spiral type points remain. Thus there exists a critical viscosity 
parameter $\alpha_{\Pi (c,i)}$ at a given ${\cal E}_{in}$ for which all the inner
saddle type sonic points are completely converted into spiral type  ones. 
In this case, the flow has no choice but to pass through the outer sonic point only.
Existence of such critical viscosities has been predicted in C90ab and C96a --- below
we compute their values exactly as a function of the inflow parameters.
This behaviour also hints  at the conclusion that the parameter space for the existance of a transonic 
flow may shrink with the increase of viscosity. 

\begin {figure}
\vbox{
\centerline{
\psfig{figure=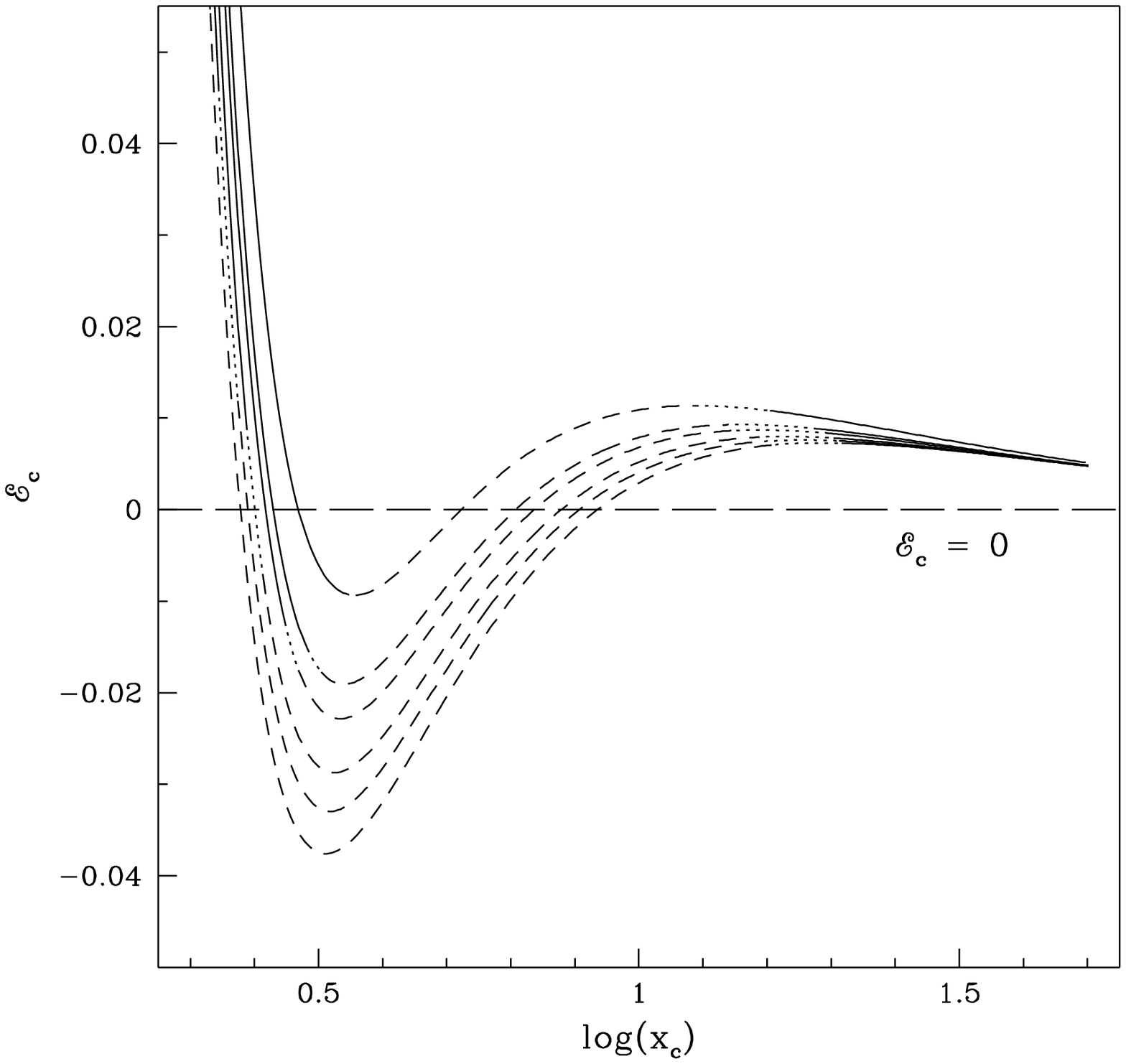,height=10truecm,width=8truecm}}}
\noindent{\small {\bf Fig. 1c:}
Variation of the specific energy at the inner sonic point as a function of the viscosity parameter
$\alpha_\Pi$. From the uppermost curve to the lowermost curve, $\alpha_\Pi = 0,\ 0.25,\ 0.35,\  
0.5,\  0.6$ and $0.7$ respectively. Other notations are the same as in Fig. 1a.}
\end{figure}

We continue our investigation of the transonic nature of the flow
and replotted a Figure similar to Fig. 1a but we increase $\alpha_\Pi$ 
gradually while keeping the specific angular momentum at the inner sonic point to
be fixed $(\lambda_c = 1.65)$. Values of $\alpha_\Pi$ are, from the top to the 
bottom curve, $\alpha_\Pi = 0,\ 0.25,\ 0.35,\  0.5,\  0.6$ and $0.7$
respectively. Solid, dotted and short-dashed lines
represent the saddle type, nodal type and the spiral type sonic points respectively. 
Long-dashed line  separates the positive and negative energy regions in the
${\cal E}_c - x_c$ plane. 
Notice that, for increasing $\alpha_\Pi$, saddle type sonic points
are gradually replaced by the nodal and spiral type sonic 
points: outer saddle type sonic points recede further away and the inner saddle 
sonic points proceed toward the black hole horizon. For $\alpha_\Pi = 0.7$, 
the inner saddle type sonic points completely disappear and become spiral type. This 
behaviour points to a critical value of viscosity parameter
($\alpha_{\Pi, c}$) which separates the accretion flow from the multiple 
sonic point regime to the single sonic point regime at a given $\lambda_c$.
It is also clear that at the same sonic point, specific energy steadily decreases for 
increasing $\alpha_\Pi$. This is because, when $\alpha_\Pi$ is increased, the accreting
matter tends to become a Keplerian disk closer to the black hole
and becomes more strongly bound with lower energy.
Note that the energy at the outer sonic point remains 
always positive for the all initial parameters. 

\begin {figure}
\vbox{
\centerline{
\psfig{figure=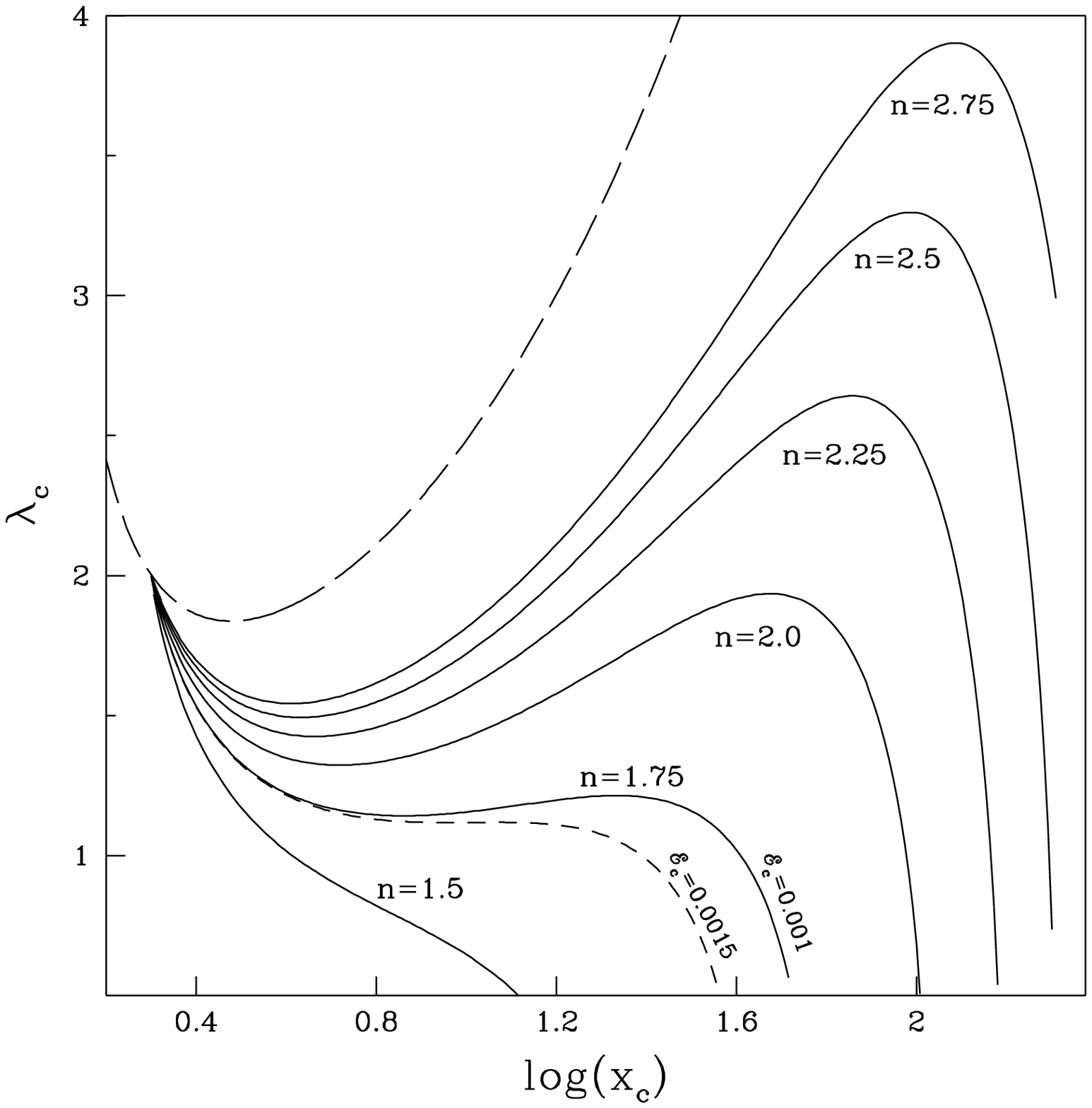,height=10truecm,width=8truecm}}}
\noindent{\small {\bf Fig. 1d:}
Variation of the specific angular momentum at the inner sonic point as a function of the
polytropic index $n$ (marked on each curve). 
Generally,  the number of sonic points decreases by decreasing $n$. Specific energy has been kept fixed at
$0.001$ except for the dashed curve where it is $0.0015$ to show that for a given polytropic
index, number of sonic points increases with decreasing energy. }
\end{figure}

In our final study of the nature of the sonic points, we chose $n$, the polytropic index,
to be our free parameter. For a very relativistic flow, or a radiation dominated flow,
$n=3$, but for a mono-atomic, non-relativistic gas, $n=3/2$. In
Fig. 1d, we show the variation of $\lambda_c$  with sonic point location $x_c$.
We keep the specific energy at the sonic point to be $0.001$. The long-dashed curve
is the Keplerian distribution as before. We note that with the increase of the adiabatic index
$\gamma$, i.e., decrease of the polytropic index $n$, the number of sonic points
decreases from three to one. In this example, 
in the extreme non-relativistic regime ($n = 1.5$) the accretion flow
has a single saddle type sonic point for any specific angular momentum. 
For the same energy, for $n=1.75$ there are three sonic points,
indicating that a standing or an oscillating shock in the flow may be possible.
In this Figure, we also show that for a given $n$ (such as  for $n=1.75$) if we increase 
energy at the sonic point $({\cal E}_c=0.0015)$, multiple sonic points
disappear and a single sonic point forms.
This indicates that there must be a critical value of ${\cal E}_c$ associated
with each $n$ above which multiple sonic points do not exist when all other
parameters are kept fixed.

\begin {figure}
\vbox{
\vskip 5.0cm
\hskip -5.0cm
\centerline{
\psfig{figure=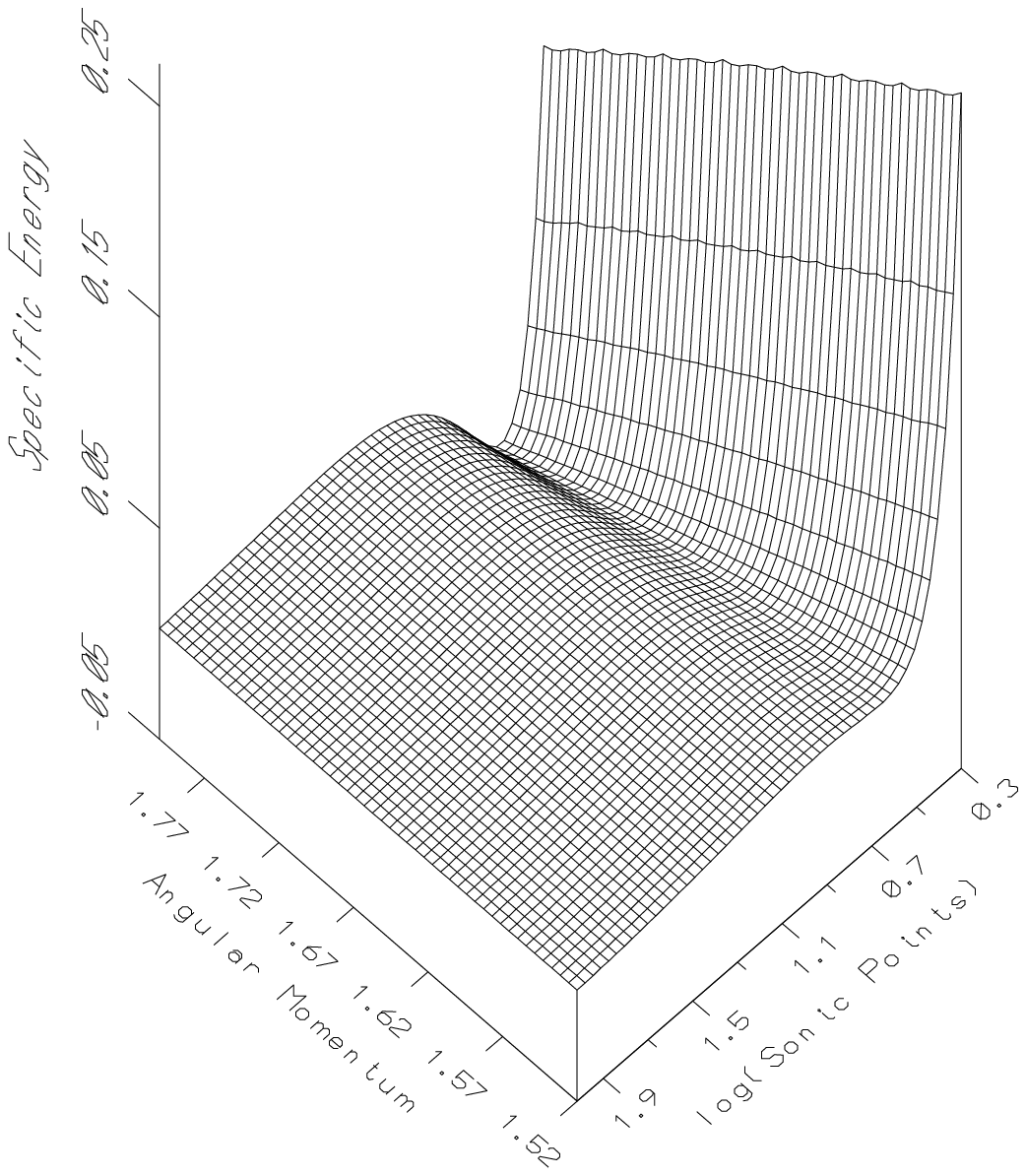,height=3truecm,width=2.0truecm}}}
\noindent{\small {\bf Fig. 2:}
Gradual change in the number of physical sonic points is easily
seen in this three-dimensional view of $F({\cal E}_c, \lambda_c, x_c)=0$ (Eq. 9) surface.
At high angular momenta, there are three sonic points, but they merge to become 
one at lower angular momenta. $\alpha_\Pi=0.01$ has been chosen.
}
\end{figure}

The general behaviour of the flow at the sonic point is 
best seen in Fig. 2, where we depict the surface $F({\cal E}_c, \lambda_c, x_c)=0$ (Eq. 9)
for $\alpha_\pi = 0.01$. Sonic points $x_c$ are plotted along X-axis in the logarithmic scale, 
$\lambda_c$ is plotted along the Y-axis and ${\cal E}_c$ is plotted along the
Z-axis. At high angular momenta, there are three sonic points, but 
they merge to become one at lower angular momenta. Below a 
critical value $\lambda_c$, the flow does not have more than 
one sonic point.

\section{Global Solution Topology}

Basic criteria for studying shock properties is that the accretion flow must
have a multiple saddle type sonic points and the shock should 
join two solutions --- one passing through the outer sonic point and the
other passing through the inner sonic point. The solution topologies have been 
already discussed in C96a. The current paper studies in great detail the
topologies associated with $f=1$ of C96a. In particular, we show 
below new pathways through which topologies may vary. 

\begin {figure}
\vbox{
\centerline{
\psfig{figure=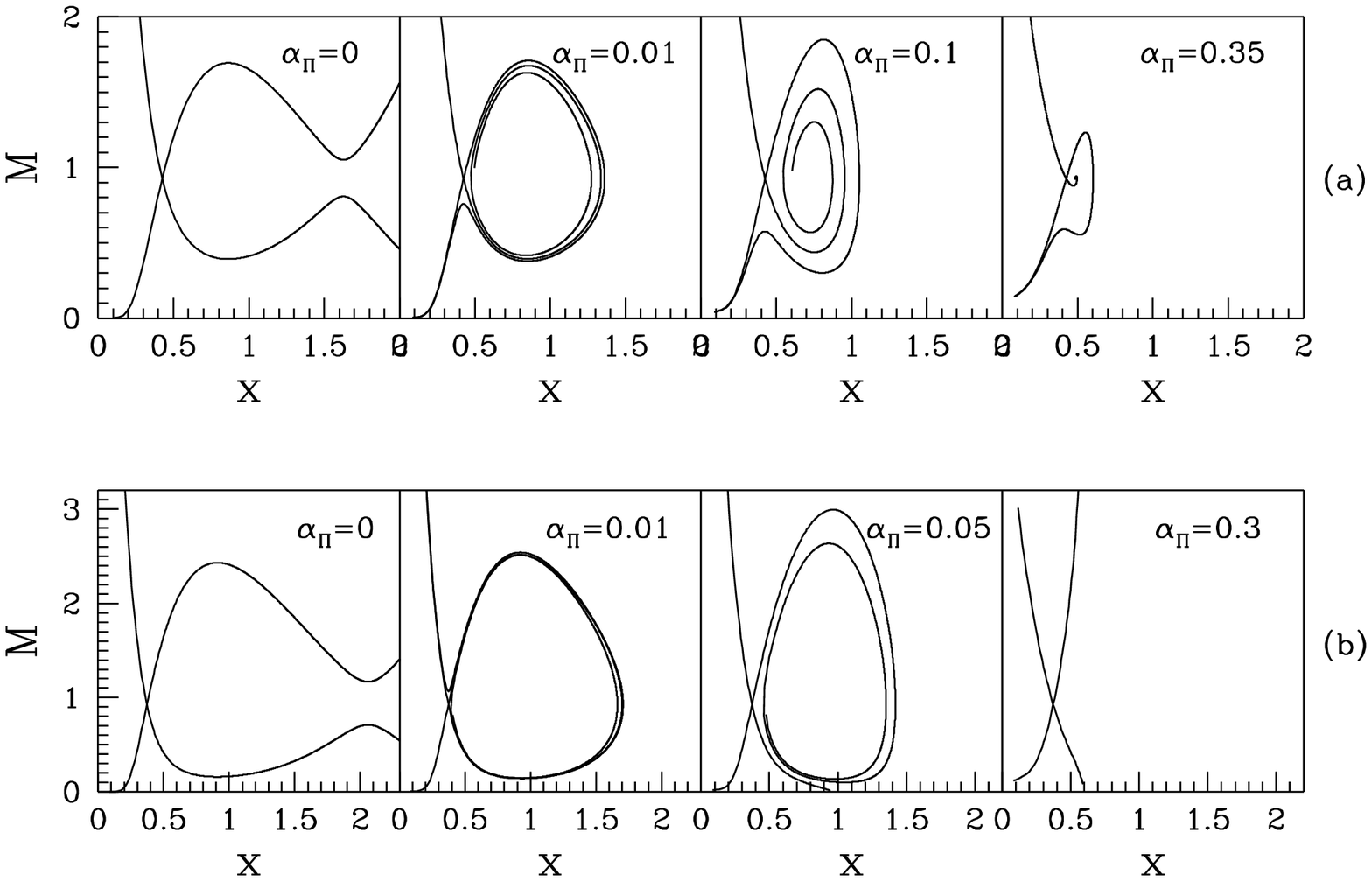,height=10truecm,width=10truecm}}}
\noindent{\small {\bf Fig. 3(a-b):}
Variation of global solution topologies of the viscous accretion flow
around black holes.  In (a), drawn for $\lambda(x_{in}=2.665)=1.68$, 
four panels show how open topology at lower viscosity becomes closed at
higher viscosity. In (b), drawn for $\lambda(x_{in}=2.359)=1.78$, 
closed topology opens up again. }
\end{figure}

In Fig. 3(a-b), we have shown how the flow topologies change with the
viscosity parameter $\alpha_\Pi$ and the
specific angular momentum $\lambda_{in}$ at the inner sonic point $x_{in}$. 
Two distinct behaviours have been highlighted here: one at a low angular 
momentum regime (Fig. 3a) and other at a high angular momentum regime (Fig. 3b).
In Fig. 3a, we keep the inner sonic point  fixed at $x_{in}=2.665$ and the specific
angular momentum at this point is $\lambda_{in}=1.68$.
At low angular momentum and without viscosity (the box at extreme left in Fig. 3a) the
subsonic flow enters into the black hole after passing through the inner sonic point. 
In the second box, viscosity is slightly higher and topologies are closed 
for the same inner sonic point. So, for a given set of parameter, there must 
be a critical viscosity parameter $(\alpha_{\Pi c})$ for which 
open topologies become closed ones.  We will
discuss critical viscosity rigorously in the \S 7. Accretion with
parameters causing this kind of topology never joins with any Keplerian disk unless
a shock is formed (This will be shown below.). When a standing shock formation is 
not possible, an accretion flow passes through the outer sonic point 
directly before falling in to the black hole. For a further increase of 
$\alpha_\Pi$ (next two boxes) closed topology shrinks gradually and finally disappears 
leaving behind only the outer sonic point (Bondi Type). This is directly analogous to 
the shrinking of the phase space of a simple harmonic oscillator in presence of 
damping (C90a). These solutions are basically the same as the $f=1$ case of Fig. 2a of C96a. 

In Fig. 3b, where solutions are plotted with a higher specific angular momentum at the 
inner sonic point ($x_{in}=2.359$ and $\lambda_{in}=1.78$), explanations of first 
and second box is similar to earlier one (Fig. 3a), but in the third box ($\alpha_\Pi=0.05$), 
accretion flow topology reverts its direction of spiraling
and the flow can join with a Keplerian disk very close to a black hole.
All the differences between this two
figures (Fig. 3a. and Fig. 3b) is mainly due to the difference of specific 
angular momentum at the sonic point rather than the change of sonic point
locations. In Fig. 11 (below), we will show that the nature of the accretion flow topologies 
have strong dependence on angular momentum at the sonic point.
For a higher $\alpha_\Pi$ (next two boxes) Keplerian disk comes even closer to the black
hole and topologies passing through the inner sonic point becomes Bondi type.
We suspect that two limits of viscosity parameters would cause an oscillation
of the inner part of the Keplerian disk, but we cannot be certain about it
without a time dependent numerical simulation.  

\begin {figure}
\vbox{
\centerline{
\psfig{figure=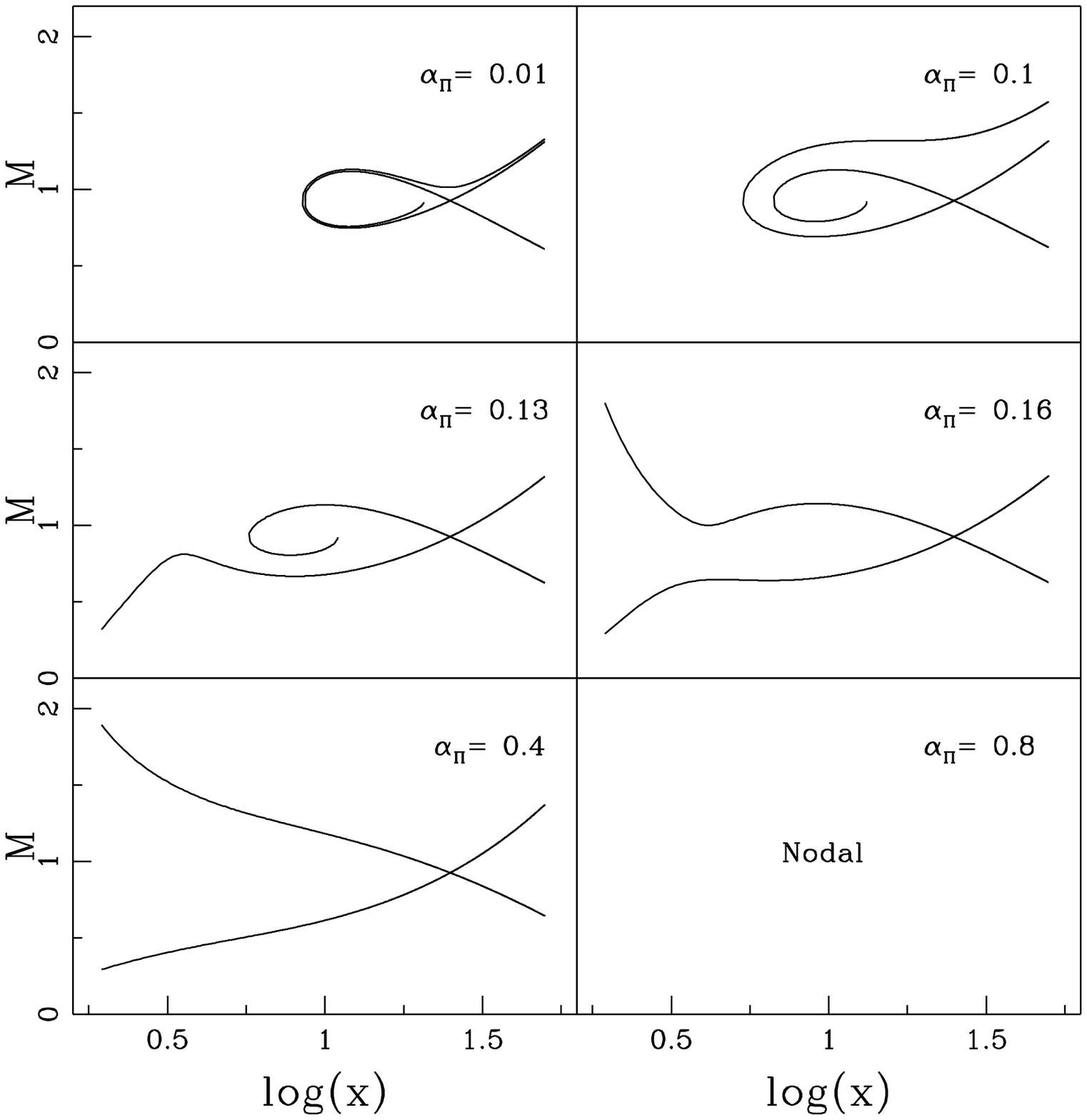,height=10truecm,width=10truecm}}}
\noindent{\small {\bf Fig. 4:}
Variation of global solution topologies of the viscous accretion flow
when the outer sonic point is kept fixed at $x_{out}=30r_g$ while the viscosity parameter is varied.
The closed topologies (with one saddle type and a spiral type at the center) 
at lower $\alpha_\Pi$ become open. Eventually, the saddle type also disappears
to produce a nodal type sonic point. }
\end{figure}

In order to show that flow topologies often can take new pathways than
what was already known (C96a), we draw in
Fig. 4, we have plotted solution topologies passing through the 
outer sonic point chosen at $x_{out}=25r_g$ and specific angular momentum $\lambda (x_{out})$ is $1.8$.
The viscosity parameter is varied (marked on each box). 
For a lower $\alpha_\Pi$, the topologies are closed as in Fig. 4 of C96a and the flow
having this topology cannot be transonic anyway. When $\alpha_\Pi$
is increased, closed topologies gradually open up (unlike C96a
where $x_{out}=35$ was chosen and the opening of topologies did not occur) and if shock 
condition is satisfied, the accretion flow passing through the outer 
sonic point jumps in to the subsonic branch and
go through the inner sonic point before entering into 
the black hole. For higher $\alpha_\Pi$, the same outer sonic point no 
longer remains saddle type. First it becomes of nodal type and then it becomes of
spiral type (Fig 1a). Considering that the outer sonic point recedes
farther away with the increase of $\alpha_\Pi$, this behaviour is not surprising.
This, together with Fig. 4 of C96a show that there could be more than one way 
of reaching nodal topology.

\section{Classification of the Parameter Space}

\begin {figure}
\vbox{
\centerline{
\psfig{figure=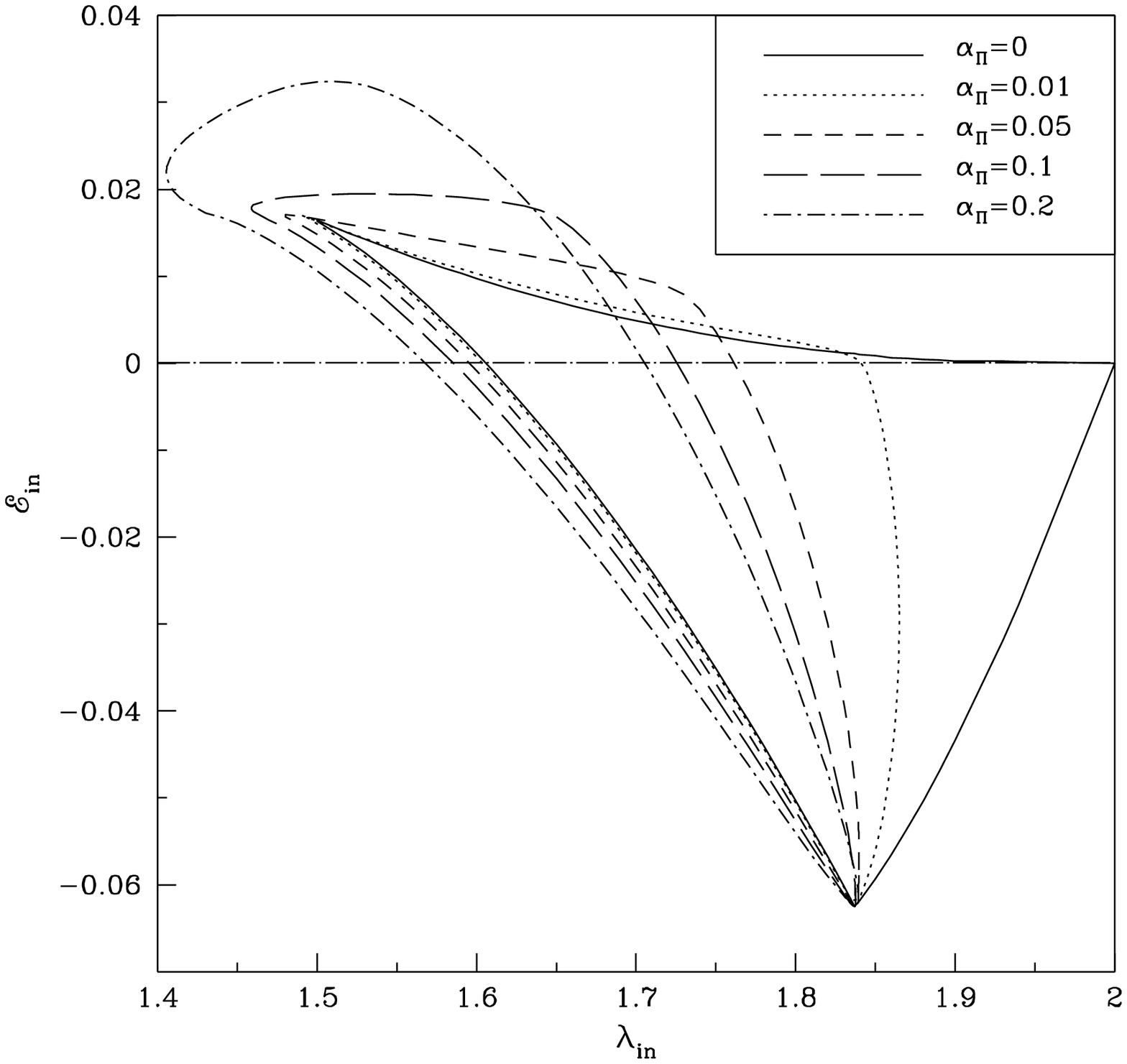,height=10truecm,width=10truecm}}}
\noindent{\small {\bf Fig. 5:}
Classification of the parameter space spanned by the specific energy 
and the specific angular momentum of the flow  at the inner sonic point.
The bounded regions drawn for different viscosity parameters (marked at the
inset)
contain allowed solutions which may pass through the inner sonic point. As viscosity
increases, the region shifts towards lower angular momentum and higher energy.}
\end {figure}

An important part of understanding a viscous flow is to classify the
parameter space as a function of the viscosity parameter. 
In Fig. 5, we have separated the parameter space for the 
accretion flow which can pass through the inner sonic point. Angular momentum
at this sonic point $(\lambda_{in})$ is varied along the X-axis and the 
corresponding specific energy at the inner sonic point $({\cal E}_{in})$ is plotted
along the Y-axis. The region bounded by a given curve contains parameter space
in which multiple sonic points are possible. For instance, for $\alpha_\Pi=0$, the 
region bounded by the solid curve is identical to the region found by C89b.
For increasing $\alpha_\Pi$, the region of multiple saddle type sonic point
is reduced near high angular momentum side while it increases in the 
lower angular momentum side. It may be recalled (C89b and Fig. 1a) that 
at low angular momentum, the number of sonic point is just one.
With the rise of $\alpha_\Pi$, the angular momentum at the sonic point
is increased, increasing the number of sonic point. 
At a higher angular momentum, the situation is just the opposite. In this case,
there are already multiple sonic points for $\alpha_\Pi\sim 0$ and 
for high enough $\alpha_\Pi$, viscosity transports angular momentum 
very rapidly causing a steep rise in angular momentum itself.
This, in turn, means that the flow can have only one saddle type sonic point in this case.

\section {Standing Shocks and Further Classification of the Parameter Space}

Though in general astrophysical contexts, shocks are ubiquitous and possibly non-stationary, 
in an accretion flow,
the location, strength and thermodynamic quantities could be quantified
very exactly by using the Rankine-Hugoniot conditions (RHCs). The study is 
similar to the study of shocks in solar winds (e.g., Holzer \& Axford, 1970) 
and white-dwarf surfaces.

The shock conditions which we employ here are the usual RHCs presented in C89b, ${\it i. e.}$,
(a) the local energy flux is continuous across the shock; (b) the mass flux is continuous 
across the shock (c) the momentum balance condition is satisfied and finally, 
(d) angular momentum should be continuous across the axisymmetric shock.

The way an accretion flow moves around a black hole, as seen from local rotating frame, is
as follows. First the flow, subsonic at a very far away, passes through the outer sonic point and becomes 
supersonic. The RHCs then decide whether a shock will be formed or not. Of course, 
our consideration of satisfying RHCs at a given location holds only if the 
shock is thin, i.e., viscosity is low. Nevertheless, we continue to use this
prescription at higher viscosities to have a first order guess of the shock location.
Similarly, we assume that there is no excess source of torque
at the shock itself, so that the angular momentum may be assumed to be continuous
across it. This condition may be violated when magnetic fields are 
present. In the presence of large scale poloidal magnetic fields, there could be magnetic torques
which could make flow angular momentum discontinuous.

\subsection{Method of calculating the shock locations}

Accretion flow after passing through outer sonic point jumps (shock) into 
subsonic branch and becomes supersonic while crossing inner sonic point before
falling into the black hole. In our present study we begin numerical 
integration from inner sonic point and proceed towards the outer edge of the 
accretion disk to look for shock location. During integration along the subsonic branch,
it is possible to calculate all the local variables $({\it i. e.} \vartheta, a, M, \rho$ 
at the post-shock region, in terms of the initial flow parameters. We calculate 
total pressure, local flow energy, specific angular momentum at shock using 
these subsonic local variables. At the shock, total pressure, local flow energy, mass accretion
rate (one of the flow parameters) and specific angular momentum are conserved. These 
conserved quantities at the shock give the other set of supersonic local 
variables for the supersonic branch. This supersonic set of local variables 
help to get outer sonic point uniquely for a accretion flow with fixed 
inner sonic point and other initial flow parameters when integration 
takes place towards the outer edge of the black hole. Thus, the accretion 
flow can be connected with both the saddle type sonic points through 
shock for dissipative system and this determines the standing shock 
location for a given set of initial parameters.

We compute the supersonic local flow variables in terms of subsonic local flow
variables in the following way:  

Our model accretion flow is in vertical equilibrium and 
total pressure of the accretion flow at any given point is given by,
$$
\Pi = W + \Sigma \vartheta^2,
\eqno(10)
$$
where, $W$ and $\Sigma$ are the vertically averaged thermal pressure
and density respectively. 

We use the mass conservation equation (Eq. 1b) in the Eq. (10) and 
calculate the sound speed $a$ in terms of radial velocity $(\vartheta)$ at the shock 
$x_s$ in the supersonic branch which is given by,
$$
a^2 = {\cal C}_1 {\cal C}_2 \vartheta - {\cal C}_2 \vartheta^2, 
\eqno(11)
$$
where, ${\cal C}_1 = \frac {4 \pi \Pi x_s}{\dot {\rm M} I_n}$ and
${\cal C}_2 = {\frac {\gamma}{g}}$. 

From the local flow energy equation, radial velocity $\vartheta$ at shock 
in the supersonic branch can be calculated using Eq. (11) and is given by,
$$
\vartheta = \frac {-\vartheta_b + 
\sqrt {\vartheta^2_b - 4 \vartheta_a \vartheta_c}}{2 \vartheta_a} ,
\eqno(12)
$$  
where, $\vartheta_a = 2 n {\cal C}_1 {\cal C}_2$, 
$\vartheta_b = -2 n {\cal C}_1 {\cal C}_2$,
$\vartheta_c =2 {\cal E} - \frac {\lambda^2(x_s)}{x^2_s} 
+ \frac {1}{(x_s-1)}$ and ${\cal E}$ is the local flow energy.

Here, total pressure and flow energy at the shock is calculated with the 
help of the subsonic flow variables. We consider only the `+' sign as we interested to get the
local flow variables in the supersonic branch. This radial velocity is used to 
get the sound speed (Eq. 11) in the supersonic branch of the flow. Now we use supersonic
flow variables $(\vartheta ~{\rm and} ~a)$ to get outer sonic point by numerical integration
and it completes the accretion flow solution having shock.

\begin {figure}
\vbox{
\centerline{
\psfig{figure=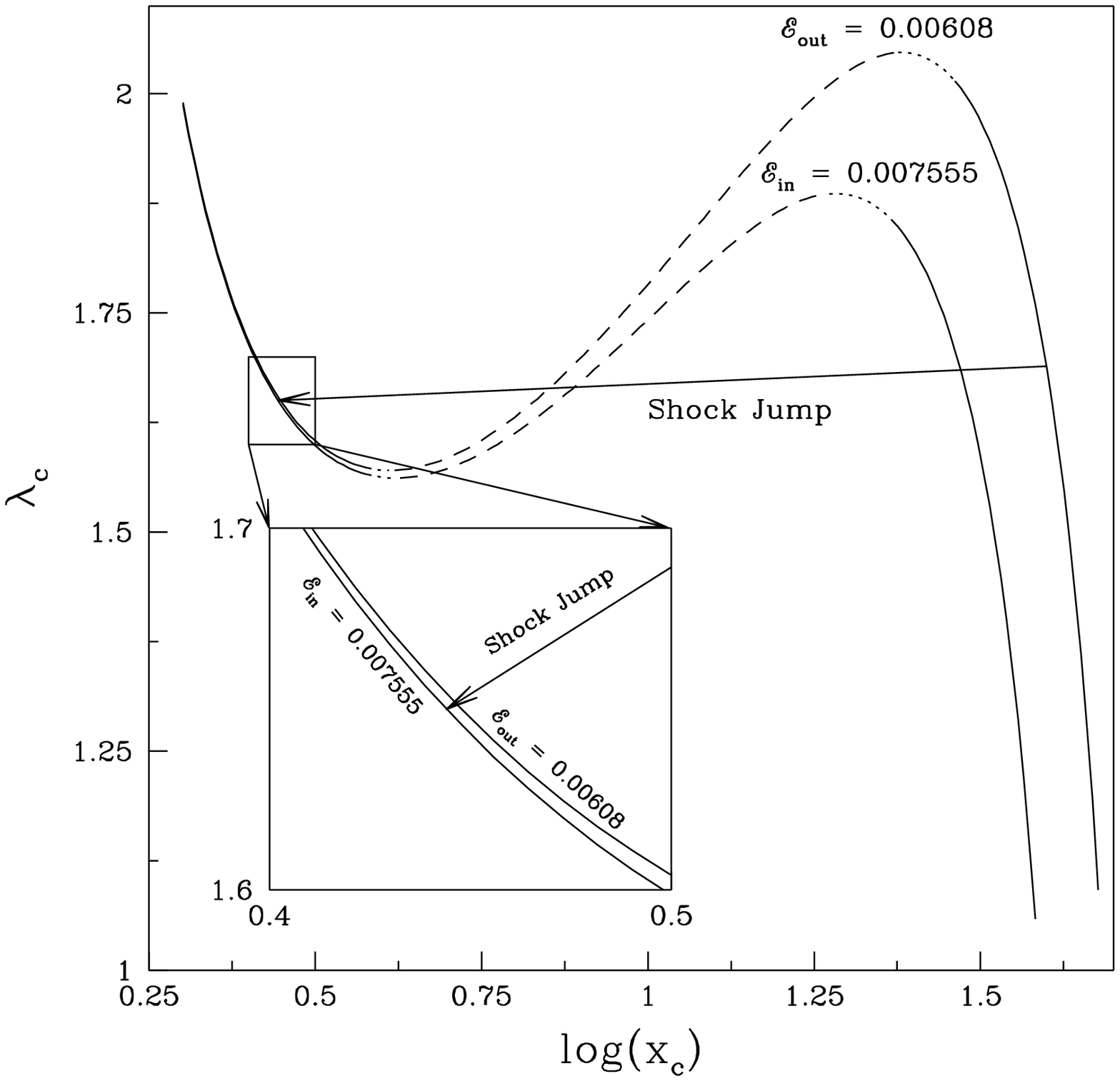,height=10truecm,width=10truecm}}}
\noindent{\small {\bf Fig. 6a:}
An example of how a standing shock might form in a viscous transonic 
flow is depicted here. Flow passing through the outer sonic point at $x_{out}=39.7$
and energy ${\cal E}_{out}=0.00608$ has a shock and passes through the 
inner sonic point at $x_{in}=2.78$ where its energy is ${\cal E}_{in}=0.007555$.
Inset shows the details.
}
\end {figure}

In Fig. 6a, we have shown how shock may be formed by joining two 
solutions, one with a lower entropy passing through the outer sonic 
point, and the other with a higher entropy passing through the inner sonic point.
Two curves are drawn for two different energies (marked).
${\cal E}_{in}$ and ${\cal E}_{out}$ are the energies at the inner and
outer sonic points for a shocked accretion flow which has a standing shock.
Due to viscous heating processes, energy is increased and 
the shock wave is formed when flow jump from the lower energy solution to higher energy solution. 
If we included only the cooling process, the situation would have been reversed.
The flow parameters are $x_{out}=39.7$, $\lambda_{in} = 1.65$, $\alpha_\Pi = 0.05$ and
$\gamma = 4/3$. The shock condition uniquely determines the 
inner sonic point, which is at $x_{in}=2.78$.
The end positions of the long arrow mark the locations of the sonic points.
In the {\it inset}, we zoom a selected region in 
the $\lambda_c-x_c$ plane to show explicitly that the 
angular momentum is indeed decreased. 

\begin {figure}
\vbox{
\centerline{
\psfig{figure=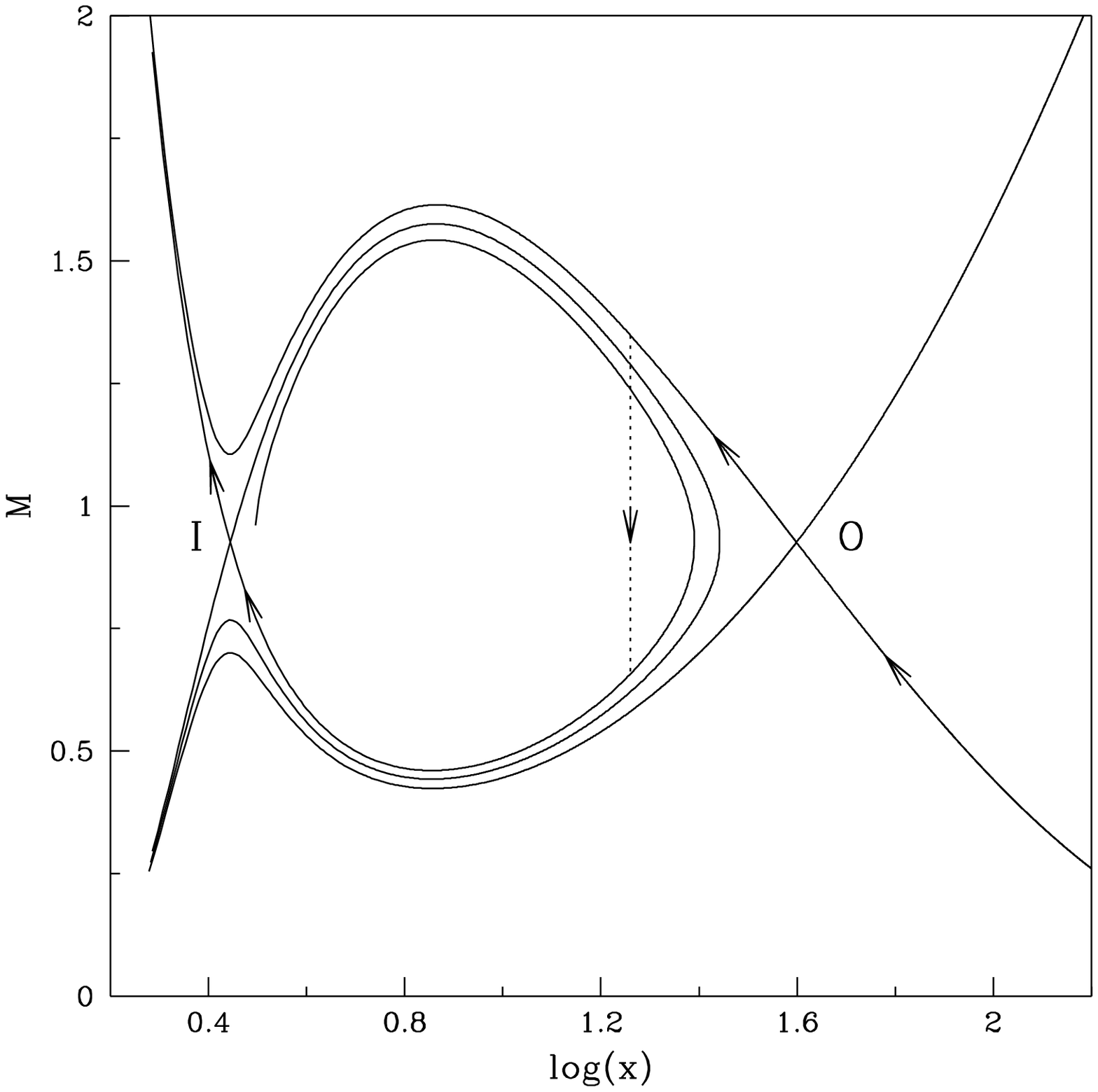,height=10truecm,width=10truecm}}}
\noindent{\small {\bf Fig. 6b:}
Actual solution topology for the case discussed in Fig. 6a is 
shown here which, along with the outer (O) and inner (I) sonic points,
also showed the shock transition at $x_s = 18.2$ (vertical dotted line).
The arrowed curve is followed by a flow while entering into the black hole. }
\end {figure}

In Fig. 6b, we present the complete solution of the flow which 
includes a standing shock in a viscous flow for the same set of parameters
used to draw Fig. 6a. Arrows indicate the direction of the accreting flow. Subsonic 
accreting flow passes through the outer sonic point (O) and becomes supersonic.
At $x_s$, shock conditions are satisfied --- the flow jumps from 
the supersonic branch to the subsonic branch and subsequently pass 
through the inner sonic point (I). In this particular case, the shock 
conditions are satisfied at $x_s = 18.2$ and shock is denoted by 
the dotted vertical line.  

\subsection{Parameter space which allows standing shocks}

\begin {figure}
\vbox{
\centerline{
\psfig{figure=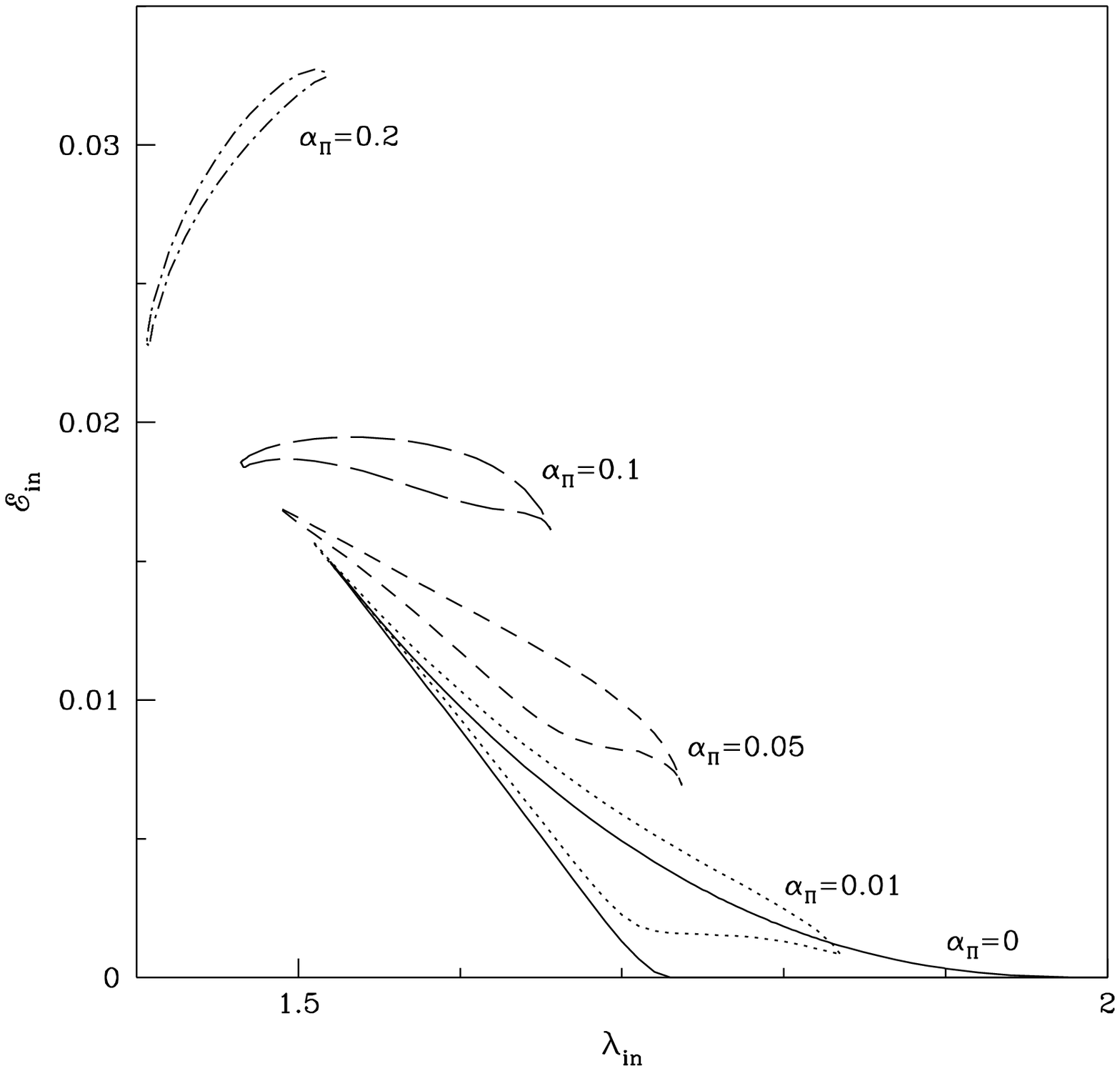,height=10truecm,width=10truecm}}}
\noindent{\small {\bf Fig. 7:}
Variation of the region of the parameter space which forms a standing shock
as a function of the viscosity parameter $\alpha_\Pi$. The region
shrinks with the increase of viscosity parameter. }
\end {figure}
In Fig. 5, we have already classified the parameter space in terms 
of the number of sonic points in the flow. 
Presently, in Fig. 7, we concentrate on the region which allows only standing shocks
in a viscous flow.  The viscosity parameters are marked.
The region marked $\alpha_\Pi=0$ coincides with that in C89b and
in Chakrabarti (1996b, 1998; hereafter C96b and C98 respectively) when
appropriate models are considered. Compared to the inviscid case, the effective 
region of the parameter space shrinks in the high angular momentum 
side when the viscosity is increased. The situation is exactly the
opposite at the lower angular momentum side. We observe that even 
at angular momentum as low as $1.4$, standing shocks could be 
formed if the viscosity is high enough. Above a critical viscosity 
(which depends on other parameters as will be shown in Fig. 13 below), this region disappears completely.   

\begin {figure}
\vbox{
\centerline{
\psfig{figure=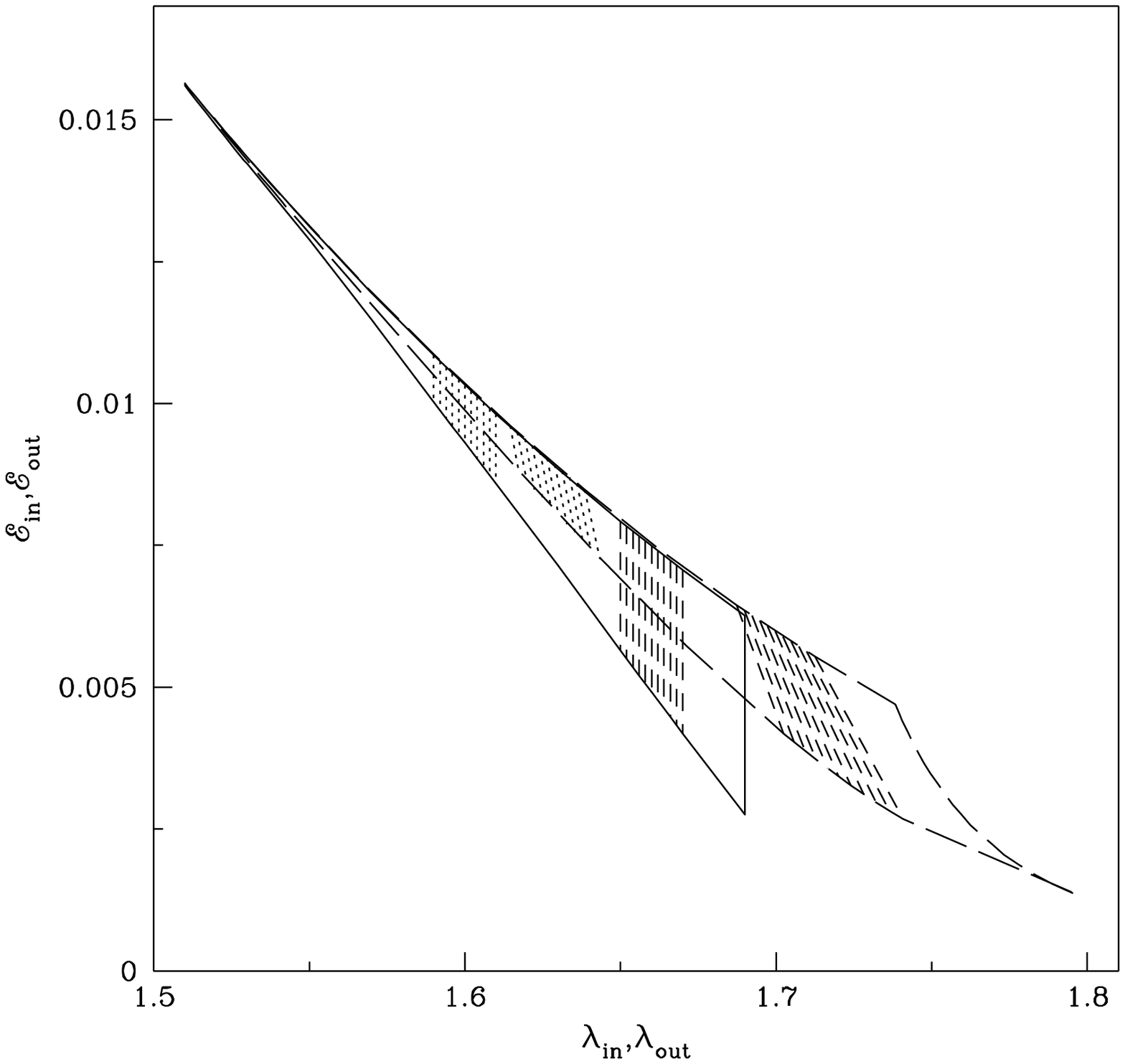,height=10truecm,width=10truecm}}}
\noindent{\small {\bf Fig. 8:}
Mapping of the parameter space of the pre-shock region (solid boundary) spanned by
$({\cal E}_{in}, \lambda_{in})$ onto the
parameter space in the post-shock region (dashed boundary) spanned by
$({\cal E}_{out}, \lambda_{out})$ in a viscous flow ($\alpha_\Pi=0.01$). }
\end {figure}

We continue our study of the parameter space which may allow
multiple sonic points. In Fig. 8, we show a curious feature: 
the mapping between the post-shock parameters and the pre-shock parameters.
We plot the region of the post-shock parameters at the inner
sonic point $({\cal E}_{in}, \lambda_{in})$ (bounded by the solid curve)
and the region of the pre-shock parameters at the outer
sonic point $({\cal E}_{out}, \lambda_{out})$ (bounded by the long dashed-curve) for a shock which
is determined through RHCs for $\alpha_\Pi = 0.01$. For each and every point 
in the pre-shock parameter space region, there 
exists a point in the post-shock parameter space region and therefore we have a complete solution.
For definitiveness, we also show vertical dashed and dotted lines in the two 
different angular momentum range in the post-shock parameter space 
region in which $\lambda_{in}$ is kept fixed but ${\cal E}_{in}$ is varied. 
The corresponding pre-shock parameters form a curve, indicating that 
both the angular momentum and the energy had to be adjusted to get the 
self-consistent solution. In a non-dissipative flow, there is no variation of energy 
and angular momentum in the accretion flow. As a result, 
both the inner and the outer sonic point parameter spaces merge (C89b, C98).

\subsection{Parameter space which may allow oscillating shocks}

An important role that is played by oscillating shocks is to produce the 
so-called quasi-periodic variations of X-ray intensity from 
galactic black hole candidates. In the inviscid flow, the 
region of the parameter space which produced multiple sonic points, 
and yet, RHCs were not satisfied was important for 
such type of oscillating shocks (Ryu, Chakrabarti \& Molteni, 1997). Here, the
winds are also produced sporadically from the post-shock region. In presence of 
cooling, especially when the cooling time scale roughly agree with the 
infall time-scale, otherwise standing shocks the shocks may oscillate 
(Molteni, Sponholz and Chakrabarti, 1996). 

It is to be noted that it is, in a general flow, very difficult to divide the
parameter space in terms of whether the shock will exist or not.  This is 
because, when there are shocks, at least the RHCs allow us to map the
pre-shock and the post-shock flow parameters (see, Fig. 8). But when there are
no shocks, it is not straight forward to map these two sets of parameters.
Thus, one has to rely on global topological behaviour of the flow solutions
and whether they allow multiple sonic points.

\begin {figure}
\vbox{
\centerline{
\psfig{figure=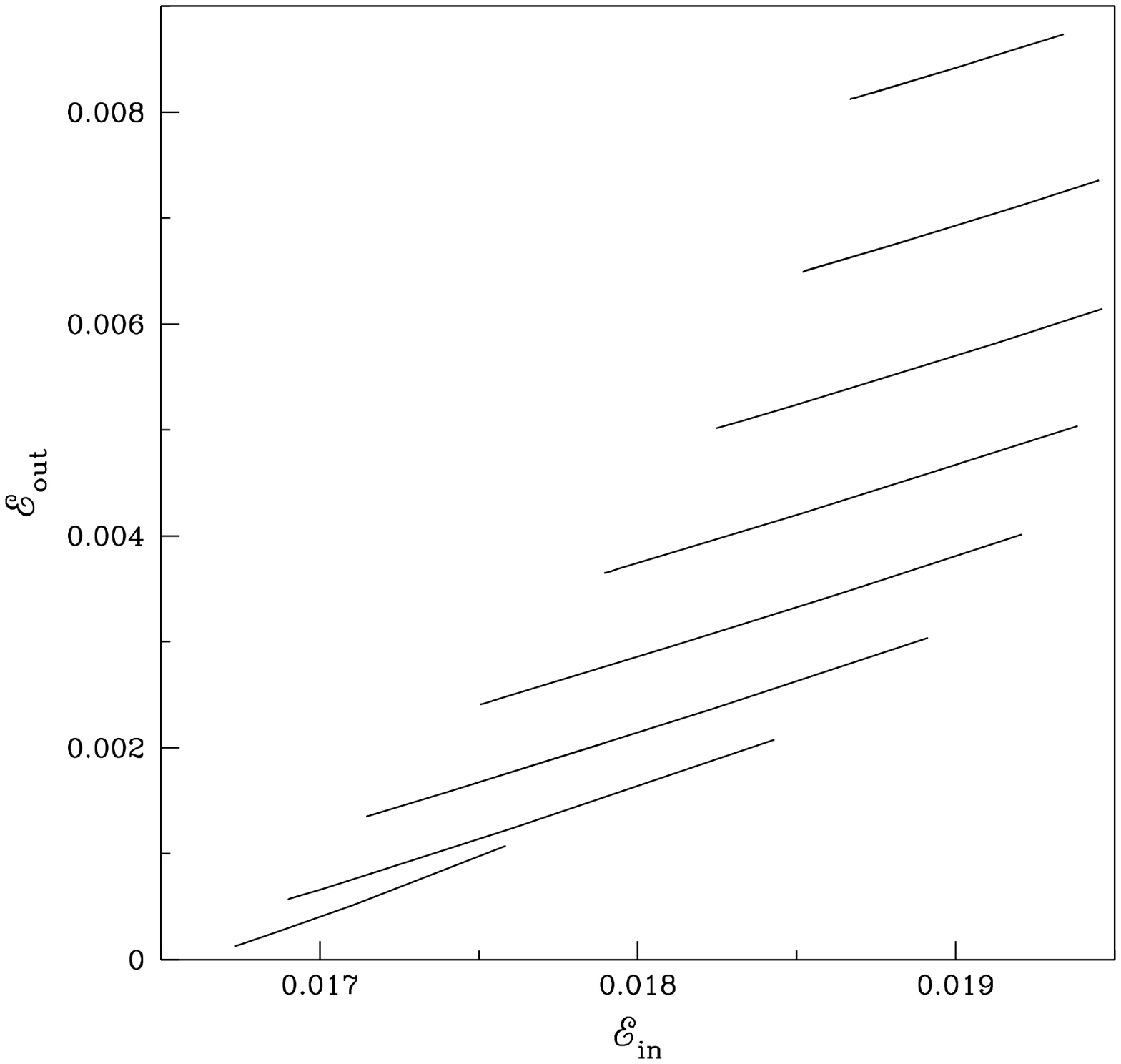,height=10truecm,width=10truecm}}}
\noindent{\small {\bf Fig. 9a:}
Example of variation of the outer sonic point energy ${\cal E}_{out}$ as a function of the
inner sonic point energy ${\cal E}_{in}$ when the flow has a shock. $\alpha_\Pi=0.1$
and $\lambda(x_{in})=1.50$ for the topmost curve.
Curves have an increment of $\Delta \lambda(x_{in})=0.02$ while going towards the bottom. }
\end {figure}

One of the criteria to use is to check which parameter space allows one to have the 
inner-sonic point energy to be larger compared with the outer sonic point energy.
For instance, in Fig. 9a, we plotted the inner sonic point energy $({\cal E}_{in})$
along X-axis and outer sonic point energy $({\cal E}_{out})$ along the
Y-axis for a set of inner sonic point angular momentum $(\lambda_{in})$ 
when accretion flows pass through shocks. It is clear that ${\cal E}_{out}$ varies 
almost linearly with ${\cal E}_{in}$ and the nature of this variation 
depends only on $\lambda_{in}$. It is not unwarranted to assume that a similar
linear variation will continue for shock-free solutions also at least if the viscosity is 
low. Thus we extrapolate this variation for the shock-free solution for the 
lower values of ${\cal E}_{in}$ till when ${\cal E}_{out}\sim 0$ for the same $\lambda_{in}$ 
keeping in mind that accretion flow topology passing through the inner sonic point
must remain closed (Fig. 3a-b). We follow this procedure to estimate the 
cut-off ${\cal E}_{in}$ for different $\lambda_{in}$
and obtain the region in the parameter space where accretion 
flow has more than one X-type sonic point.

\begin {figure}
\vbox{
\centerline{
\psfig{figure=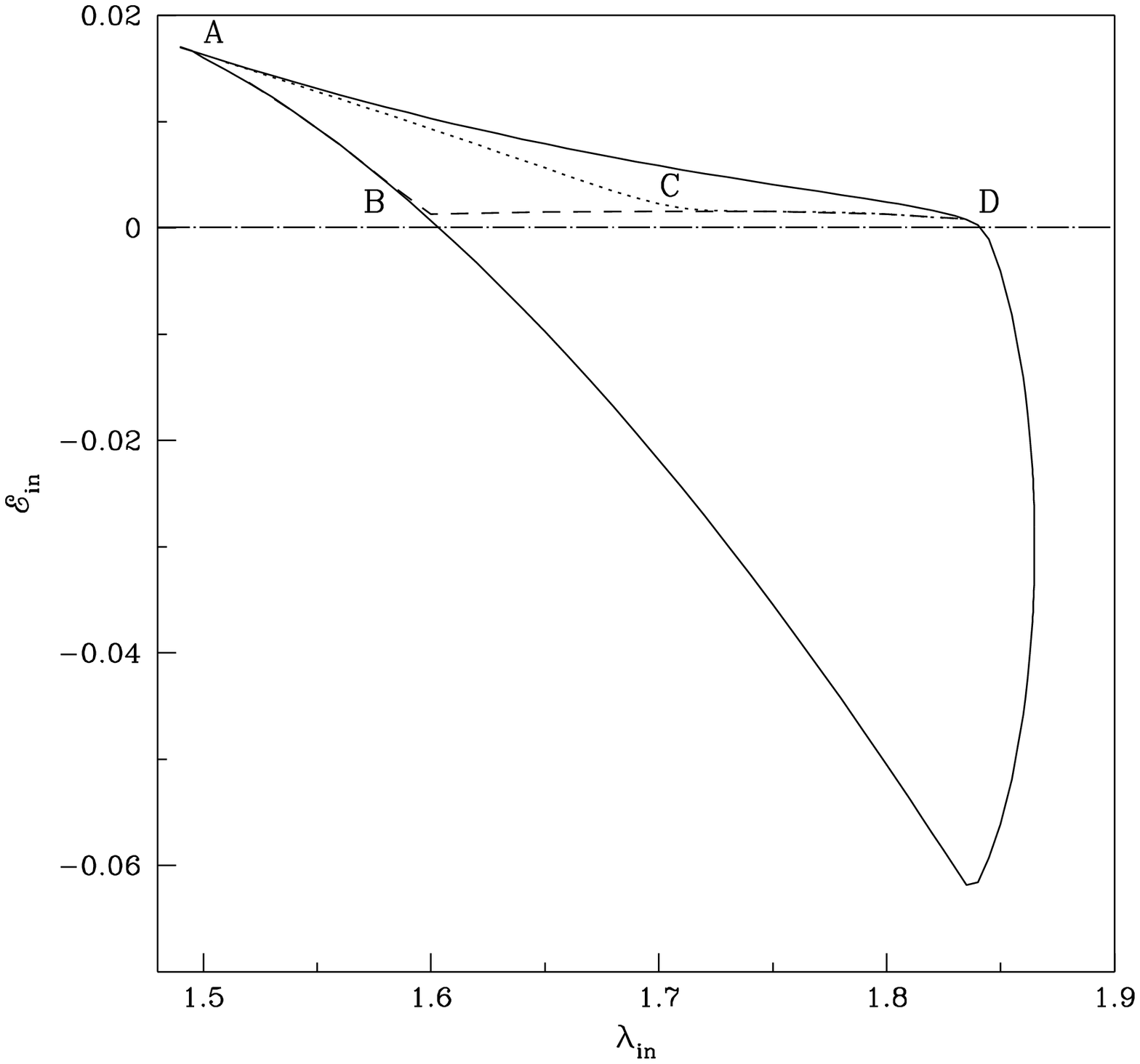,height=10truecm,width=10truecm}}}
\noindent{\small {\bf Fig. 9b:}
Division of the parameter space $({\cal E}_{in}, \lambda_{in})$ for a viscosity parameter
$(\alpha_\Pi = 0.01)$ on the basis of number of sonic points. Region separated by the 
dotted line has more than one X-type (saddle type) sonic points and flows in this region form 
standing shocks. The region surrounded by the dashed courve has more than one X-type 
sonic points but the Rankine-Hugoniot conditions are not satisfied here. }
\end {figure}

In Fig. 9b, we show the division of the parameter space $({\cal E}_{in}, 
\lambda_{in})$ for the viscosity parameter $\alpha_\Pi = 0.01$ on 
the basis of number of sonic points. Flows with parameters from the region 
ACD have more than one X-type (saddle type) sonic points and 
RHCs are also satisfied. Flows with parameters from the region ABC have
more than one X-type sonic points but RHCs are not satisfied here.
From our previous experience with non-dissipating flows, we predict that those 
solutions with multiple sonic point which do not produce standing shocks must be 
producing oscillating shocks. This region becomes bigger when the viscosity 
parameter is reduced. Rest of the parameter space gives solutions with closed
topology passing through the inner sonic point.

\subsection{Parameter space for all possible solutions}

\begin {figure}
\vbox{
\centerline{
\psfig{figure=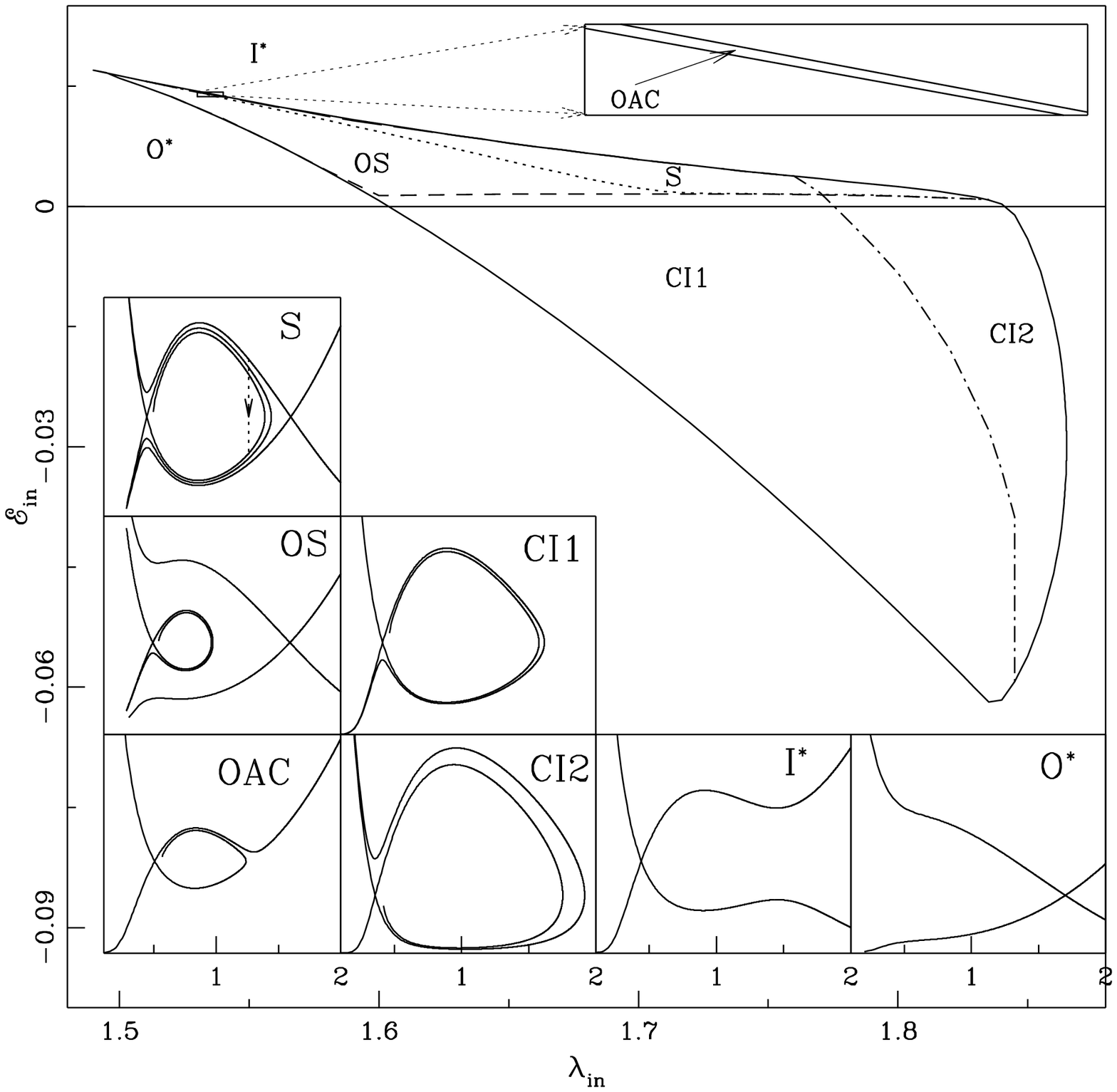,height=10truecm,width=10truecm}}}
\noindent{\small {\bf Fig. 10:}
Division of the parameter space $({\cal E}_{in}, \lambda_{in})$ (marked)
for a viscosity parameter $(\alpha_\Pi = 0.01)$ on the basis of 
solution topologies shown in boxes (marked). See text for details.
}
\end {figure}

Figure 10 shows the classification of the parameter space in the energy-angular
momentum $({\cal E}_{in}, \lambda_{in})$ plane in terms different accretion 
flow topologies (small box) for $\alpha_\Pi = 0.01$. Solid boundary separates the 
region in the parameter space for closed topologies passing through the inner 
sonic point in general. Further sub-classes are indicated by dotted, 
dashed and dot-dashed curve which classified the solution topologies depending 
on their behaviour. Examples of solution topologies with initial parameters 
taken from different regions (marked) of the parameter space are plotted in seven 
small boxes (marked). All the small boxes depict Mach number variation as a function of the
logarithmic radial distance. The box labeled S shows an accretion 
flow solution which passes through a shock. Dotted vertical line 
with an arrow indicates the location of the standing shock.
The solution drawn in the box marked OS is an accretion flow which has multiple 
sonic points but does not satisfy RHCs after the flow becomes supersonic. 
From our earlier experience with an inviscid flow, this topology
is expected to give rise to an oscillating shock solution. The box marked OAC shows a
new type of solution topology having multiple sonic points. One branch of the topology 
is closed and the other branch is open. This kind of solution is available 
in a small region of the parameter space shown in the inset on the upper-right 
corner. Solutions inside the CI1 box have closed topology (inner spiral going anti-clockwise)
having only one saddle type sonic point
and this kind of solution belongs to a large region of the parameter space
with relatively lower angular momentum region. The box CI2 shows a similar 
result as CI1 but here the nature of the topology is different (inner spiral going clockwise) 
and this type of solution exists at higher angular momentum region.
The box labeled ${\rm I}^*$ represents accretion flow solution
which only passes through the inner sonic point. This solution could be for an accretion
or wind and the initial parameters for this type of topology belongs to the
region indicated by ${\rm I}^*$ in the parameter space. Topology with parameters 
taken from ${\rm O}^*$ region of the parameter space
is also plotted in the box marked $O^*$. An accretion flow solution with these parameters
passes only through the outer sonic before falling into the black hole (similar to a Bondi flow).

\begin {figure}
\vbox{
\vskip 1.0cm
\centerline{
\psfig{figure=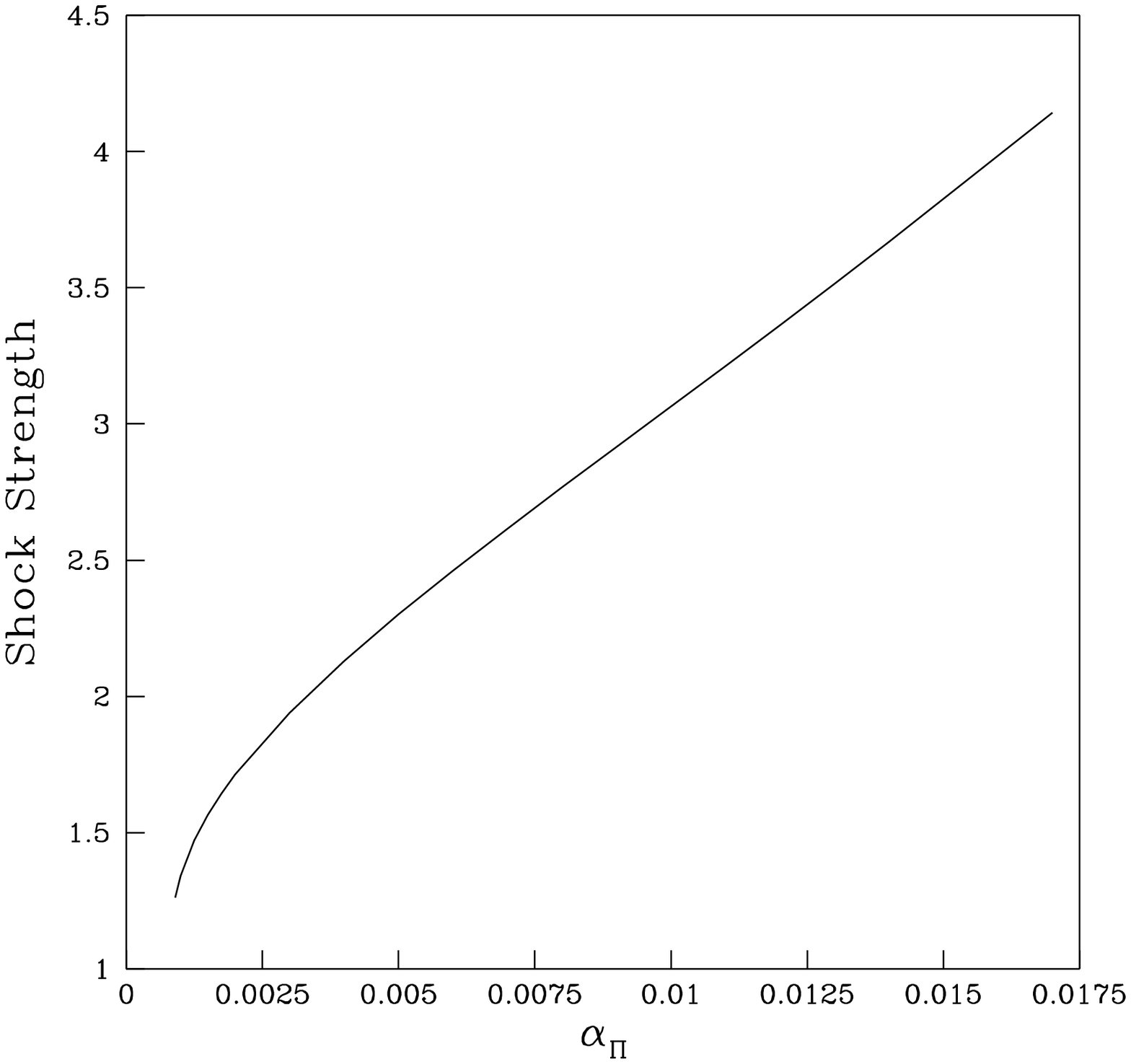,height=8truecm,width=8truecm}}}
\noindent{\small {\bf Fig. 11:}
Variation of the ratio of the pre-shock to post-shock Mach numbers
as a function of viscosity parameters for a fixed set of 
initial conditions ($x_{in} = 2.795$ and $\lambda_{in} = 1.65$). 
The shock disappears beyond the critical parameter $\alpha_\Pi \sim 0.017$.
}
\end {figure}

Since the strength of the shock determines the jump in temperature and density,
it may be worthwhile to study the shock strength. We define this as the
ratio of the pre-shock Mach number to the post-shock Mach number.
As an example, in Fig. 11, we show the variation of the shock strength 
as a function of viscosity parameter $\alpha_\Pi$. This 
Figure is drawn for $x_{in} = 2.795$ and $\lambda_{in} = 1.65$. 
For lower viscosity limit, the strength of the shock is weak. It increases smoothly 
with the gradual increase of viscosity and there is a cut off at a critical 
viscosity limit where the shock disappears.

\section{Dependence of the Critical Viscosity Parameter}

In our earlier discussion, we already hinted that there must be a 
critical viscosity parameter for which the flow topology must change
its nature from an open topology to a closed topology. Presently, we
quantify the nature of this critical viscosity. 
Indeed, we find that there are in effect two critical viscosity parameters: 
one at the boundary which separates the closed topology from the open topology
while the other splits region of closed topology in terms of 
whether shocks can form or not. Not surprisingly,
these are inflow parameter dependent, and thus do not have  universal values.
Nevertheless, these are useful, since they give us insights 
into the cases in which shocks may be possible.

\begin {figure}
\vbox{
\centerline{
\psfig{figure=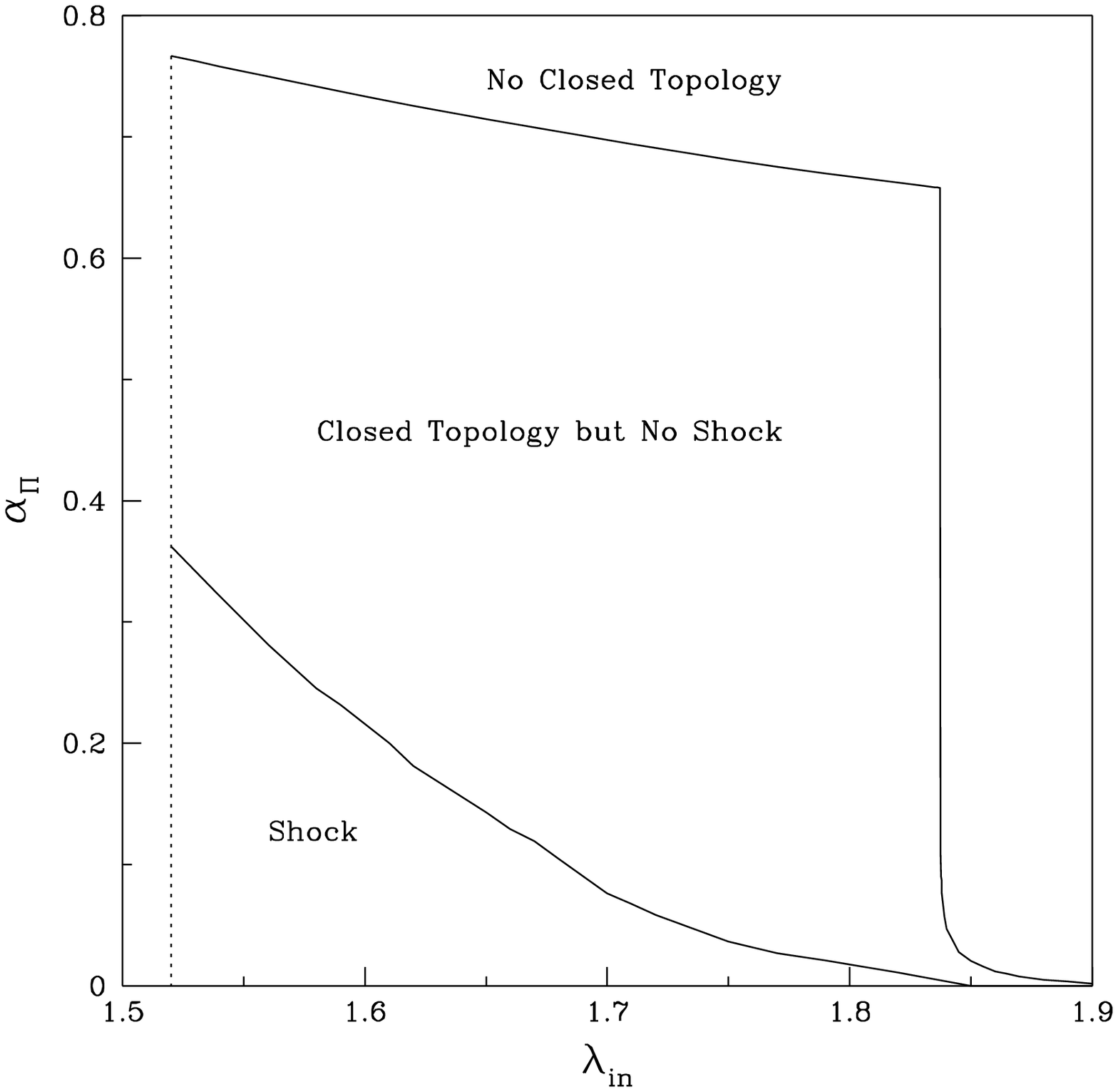,height=10truecm,width=10truecm}}}
\noindent{\small {\bf Fig. 12:}
Critical viscosities separating standing from oscillating shocks
and closed topologies from open topologies.
}
\end {figure}

Figure 12 shows the variation of the critical viscosity parameters 
with the angular momentum at the inner sonic point. Different
regions are marked. We note that 
for higher viscosity parameters $(\alpha_\pi)$, shocks are formed
in the lower angular momentum domain. As the angular momentum is increased, 
the shock disappears. This was also expected from our discussion of Fig. 1c.

\begin {figure}
\vbox{
\centerline{
\psfig{figure=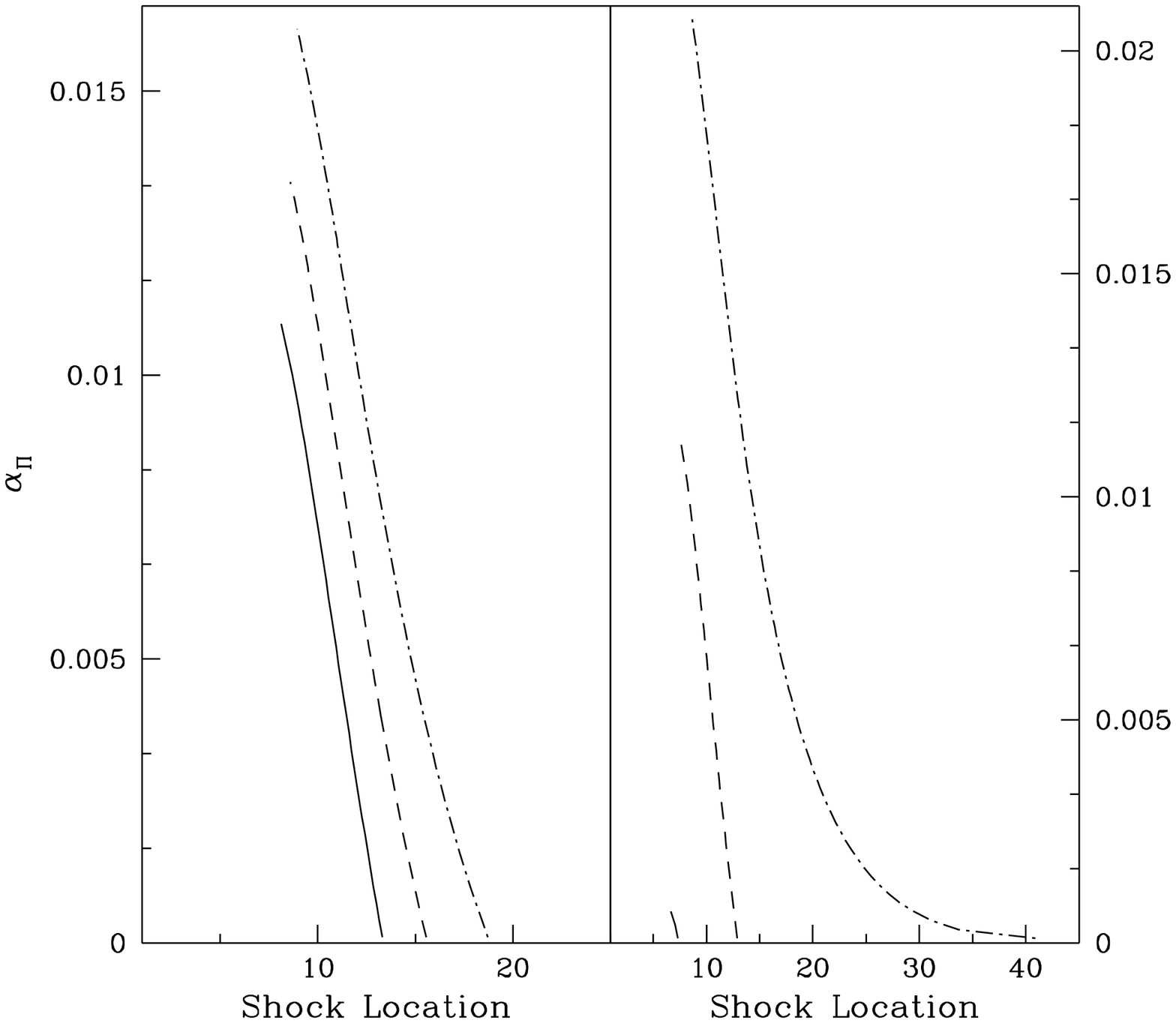,height=10truecm,width=10truecm}}}
\noindent{\small {\bf Fig. 13(a-b):}
Variation of shock location with viscosity parameter and (a) inner
sonic point and (b) specific angular momentum. Shock location always
decreases with increase of viscosity till the critical 
viscosity parameter is reached beyond which the shock ceases to exist. }
\end {figure}

\section{Dependence of the Shock Location on Viscosity Parameter}

In our study of shock properties, we have already mentioned that 
the shocks disappear when viscosity is more than a critical value.
In Fig. 13 (a-b), we show how the shock location depends on the 
viscosity parameter when the other two free parameters, i.e.,
the shock location ($x_{in}$) and angular momentum ($\lambda_{in}$)
are kept fixed. In Fig. 13a,  variation with the inner sonic point is shown
when the angular momentum is kept fixed, while in Fig. 13b, the 
variation with the angular momentum is shown keeping the inner sonic point
fixed. In {\it all} the cases, the shock location is {\it reduced} with
the increase in viscosity parameter till the critical
viscosity parameter is reached beyond which the shock ceases to
exist. This is significant because in an accretion flow,
when viscosity is increased, the accretion rate is also increased and a black hole
candidate goes from spectrally hard to spectrally soft state (CT95).
Thus if the shock oscillation is indeed the cause of quasi-periodic 
oscillations (QPOs), then the frequency should increase
with the accretion rate and finally as the shock ceases to exist, the QPOs also
should disappear in softer states. Observation of such features could be used 
to verify if the shock oscillations may be the prime cause of the QPOs
in black hole candidates.

\section{Concluding Remarks}

In this paper, we have extended our earlier results of the study of 
shock formation to include a very difficult yet more realistic case of viscous
polytropic flows. Some of the results have been touched upon in C96a
but the new results in our work include a
detailed study of the parameter space in which shocks form even in presence of
viscosity. We found a large number of important results: (a) That there 
exists two critical viscosity parameters which separate the region of the parameter space
in three parts -- 1) in which the flow has a Bondi-type single sonic point; 2)
in which there are three sonic points but no Rankine-Hugoniot relations are
satisfied and 3) when Rankine-Hugoniot relations are satisfied. These critical viscosity parameters
decrease with the increase of the specific angular momentum of the flow at the 
inner sonic point. (b) That at high viscosities, standing and oscillating shocks may form if the
flow has very little angular momentum at the inner sonic point, while
at low viscosities the situation is exactly the opposite. It is widely believed
that accreting matter on galactic and extra-galactic black holes 
could be of very low angular momentum, especially when the central compact object
is accreting winds from the nearby star or stars. This brings out the
possibility that shocks may be active ingredients of an accretion flow. Our results,
with a very plausible accretion flow models, indicate that the 
standing and oscillating shocks are produced even for large viscosity parameters. (c)
That the shock location is reduced 
with enhancement of viscosity parameter. This, coupled to earlier 
results (Chakrabarti \& Manickam, 2000) that the infall time
is proportional to the period of quasi-periodic oscillation (QPO) 
of X-rays from black holes imply that the QPO frequency should increase
as the viscosity is increased. This is consistent with the observational
findings that QPO frequency is increased as the spectral slope softens,
widely known to be due to increase in viscosity and accretion rate.

One of the questions we have not addressed here is the stability properties 
of these shocks. A number of authors pointed out that while the shocks
are stable, they should undergo oscillations, either radially, or vertically, or
non-axisymmetrically (Molteni, Toth \& Kuznetsov, 1999; Gu \& Foglizzo, 2003). 
We anticipate that our shock solutions in viscous flows would suffer
similar types of oscillations, especially when the viscosity is low. In particular, Gu 
\& Foglizzo (2003) while studying shocks in inviscid, isothermal flows
found such instability and interpreted as due to cycles of acoustic waves between their co-rotation
radius and the shock. In their interpretation this could be a form of Papaloizou-Pringle
instability (Papaloizou \& Pringle, 1984) which is known to destabilize 
accretion tori when the angular momentum gradient is less than a certain value.
If so, such an instability could disappear at high enough viscosity. 
This could have a bearing on the quasi-periodic oscillations of observed 
X-rays in galactic and extra-galactic black hole candidates in that QPOs 
would cease to exist above a certain frequency. The interesting aspect is 
that these so-called 'instability' only causes oscillation of shocks 
and does not destroy the shock (Molteni, Toth \& Kuznetsov, 1999).

This work is partly supported by a project (Grant No. SP/S2/K-15/2001)
funded by Department of Science and Technology (DST), Govt. of India.

{}

\end{document}